\documentclass[twocolumn,showpacs,preprintnumbers,amsmath,amssymb,prb]{revtex4-1}
\usepackage[bbgreekl]{mathbbol}
\usepackage{graphicx}
\usepackage{dcolumn}
\usepackage{bm}
\usepackage{mathtools}
\usepackage{physics}
\usepackage{hyperref}
\usepackage{color}
\usepackage{subcaption}
\usepackage[a4paper]{anysize}
\graphicspath{{fig/}{img/}}

\newcommand{\E}{{\bf E}}

\renewcommand{\j}{{\bf j}}
\renewcommand{\k}{{\bm{k}}}
\newcommand{\rr}{{\bm{r}}}

\newcommand{\q}{{\bm{q}}}

\newcommand{\ii}{{\rm i}}
\newcommand{\ee}{{\rm e}}

\def\gsim{\lower.35em\hbox{$\stackrel{\textstyle>}{\textstyle\sim}$}}
\def\lsim{\lower.35em\hbox{$\stackrel{\textstyle<}{\textstyle\sim}$}}

\begin{document}
\title{Non-local quantum effects in plasmons of graphene superlattices}

\author{Luis Brey}
\affiliation{Materials Science Factory, Instituto de Ciencia de Materiales de Madrid (CSIC), Cantoblanco, 28049 Madrid, Spain}
\author{T. Stauber}
\affiliation{Materials Science Factory, Instituto de Ciencia de Materiales de Madrid (CSIC), Cantoblanco, 28049 Madrid, Spain}
\author{L. Mart\'{\i}n-Moreno}
\affiliation{Instituto de Ciencia de Materiales de Arag\'on and Departamento de  F\'{\i}sica de la Materia Condensada, CSIC-Universidad de Zaragoza, E-50009, Zaragoza, Spain}
\author{G. G\'omez-Santos}
\affiliation{Departamento de F\'{\i}sica de la Materia Condensada, Instituto Nicol\'as Cabrera and Condensed Matter Physics Center (IFIMAC),
Universidad Aut\'onoma de Madrid, E-28049 Madrid, Spain}
\begin{abstract}
By using a non-local, quantum mechanical response function we study graphene plasmons in a one-dimensional superlattice (SL) potential $V_0 \cos G_0x$. 
The SL introduces a quantum  energy scale
$E_G \sim \hbar v_F G_0$ associated to electronic sub-band transitions.
At energies lower than $E_G$, the plasmon dispersion is highly anisotropic; 
plasmons propagate 
perpendicularly to the SL axis, but become damped by electronic transitions along the SL direction. 
These results question the validity of semiclassical approximations for describing low energy plasmons in periodic structures.
At higher energies, the dispersion becomes isotropic and Drude-like 
with effective Drude weights related to the average of the absolute value of the local chemical potential.
Full quantum mechanical treatment of the kinetic energy thus introduces non-local effects that delocalize the plasmons in the SL, making the system
behave as a meta-material even near singular points where the charge density vanishes.
\end{abstract}
\maketitle
{\it Introduction.}
Graphene plasmons\cite{Ju11,Chen12,Fei12} have attracted much attention due to their long propagation lengths,\cite{Ni18} strong confinement,\cite{Alcaraz18} and high gate and frequency tunability.\cite{nikitin-book,Koppens:2011aa,Low:2016aa,Torbatian,Garcia-de-Abajo:2014aa,Peres-book,Bonaccorso:2010aa,Farmer:2015aa,Fernandez-Dominguez:2017aa,Galiffi18,Fan19,Foerster19}  
Plasmons in pristine graphene are well-described by the random-phase-approximation,\cite{Wunsch:2006aa,Ando:2006aa,Hwang:2007aa} and in the long wavelength limit,
the dispersion  reads 
\begin{equation}
\omega _p =\sqrt{ \frac {D}{2\epsilon\epsilon _0}  q}\;, \label{Plasmon}
\end{equation}
where $D$ is the Drude weight, $q$  the wave vector and $\epsilon$  the dielectric constant of the surrounding medium. 
Eq. (\ref{Plasmon}) can be derived from a hydrodynamic approach and thus holds for all two-dimensional (2D) systems.\cite{Ando:1982aa} For graphene and also for parabolic bands,\cite{Stauber:2014aa} $D\to D_0=\frac {e^2}{\hbar^2} \frac {|E_F|} {\pi}$, thus in Dirac systems the plasmon frequency depends on the carrier density as $n^{1/4}$.\cite{Hwang:2007aa,Brey:2007aa,Horng11,Yoon14}
 
Exciting plasmons by incident radiation is not possible in pristine graphene because plasmons are strongly confined and energy and momentum conservation prevents their coupling.
The coupling can be achieved, though, by superimposing 
a superlattice (SL)  on graphene via external or thermal grating,\cite{Peres:2013aa,Alcaraz18,Yu18} patterning,\cite{Ju:2011AA, Drienovsky:2018aa,Xiong19} or growing graphene on vecinal surfaces.\cite{Celis:2018aa} 
The SL periodicity induces a folding of the plasmon dispersion, and  plasmon sub-bands appear in the SL Brillouin zone. At the center of the Brillouin zone, the second plasmon sub-band  has its origin 
in the folding of unperturbed plasmonic modes with momentum $\pm G_0$, where  $G_0$ is  the SL reciprocal lattice vector.
This plasmon, of finite energy, at the center of Brillouin zone is inside the light cone and can couple to incident light.

Graphene plasmons in a SL are conventionally discussed within a semiclassical (SC) and  local approximation that assumes that the system responds solely to local  external fields at each point in space.\cite{Peres:2012aa,Slipchenko:2013aa,silveiro:2013aa,Beckerleg:2016aa,Huidobro:2016aa,Huidobro:2016ab,nikitin-book} This approach is equivalent to assuming  
that the optical conductivity at each point in space is determined by the Fermi energy at that point which in turn is obtained from the local charge density using the Thomas-Fermi approximation, $\sigma (r)=\sigma _L(E_F[n(r)])$. In these calculations, the main effect of the optical conductivity modulation is the localization of  plasmons in the regions of smaller conductivity.
Recently, it has been estimated by SC approximations that non-local effects are important near regions in which the optical conductivity is strongly suppressed.\cite{Galiffi} However, those are the regions where SC approximations are expected to break down, making it necessary to develop a fully
quantum-mechanical   (QM) approach. 

{In this Letter, we study the collective excitations of graphene in the presence of a one dimensional superlattice potential
$V(x)$=$V_0 \cos{ G_0 x}$ within the  linear random phase approximation  that includes non-local and quantum-mechanical effects (Q-non-local). Throughout the manuscript, we contrast this approach with two widespread approximations: the local quantum mechanical approximation which reduces the full response matrix to a function (Q-local), and the semiclassical, local approximation which introduces a local Fermi energy in the response function (SC local).}

Our main results  can be summarize by the following points: i) absence of Klein tunneling for the motion of plasmons; ii) at low frequencies, plasmons moving along the 
SL  axis are damped by electronic SL sub-band  excitations; iii) near singular points, non-local effects associated to electron kinetic energy delocalizes the plasmons in the SL unit cell and make the system behave as a meta-material; iv) plasmonic excitations in modulated graphene can be related to an effective Drude weight even for vanishing charge density. 
 
{\it Hamiltonian.} The massless Dirac Hamiltonian for one valley and one spin-projection in the presence of an external potential is given by
\begin{equation}
H =\hbar v_F \left (-i \sigma _x \partial _x-i \sigma _y \partial _y \right )+ V(x) {\cal I}\;,
\label{hamiltonian}
\end{equation}
where $\sigma _x$ and $\sigma _y$ are the Pauli matrices, $v_F$ is the Fermi velocity, $\cal I$ is the identity matrix. Unless otherwise stated, results in this paper will be presented for an exemplary period of $L=2\pi /G_0=600a$ with $a$=$2.46${\r{A}}. This leads to the energy scale $E_G=\hbar v_FG_0\sim25$meV which is related to the SL interband electronic transitions. For details on the band-structure and further discussion, see the Supplementary Material (SM),\cite{SI} {in particular the inset of Fig. 1b).}

{\it Plasmonic response.} Plasmonic excitations are the response to an infinitesimal external potential with wave number $\q$. However, the induced charge density contains not only the incoming mode $\q$, but all higher harmonics $\q+G$ with $G=nG_0$. 
The SL potential defines a reduced Brillouin zone where the
plasmon sub-bands are defined. The optical conductivity in the $\nu$-direction thus becomes a matrix of the form
$\sigma^{\nu\nu}_{G,G'}(\q,\omega)$, 
and the following discussion is based on the energy loss obtained from the largest eigenvalues of the dielectric response and their respective eigenvectors. See SM for how these quantities are computed.\cite{SI}

Another quantity of interest will be the Drude weight, which is the static  limit of the reactive conductivity, and thus related to electronic intraband transitions. The conductivity has also contributions from electronic interband transitions and plasmons are often characterised by the electronic transitions they are composed of.\cite{Falkovsky07,Stauber08}
In pristine graphene, plasmons are usually Drude-like and have no contributions from the interband conductivity. 
However, a SL potential creates both low energy electronic sub-bands and  spectral weight transfer from electronic intraband to electronic interband transitions. 
This is a QM effect that is not included in SC calculations, in which the electronic sub-bands of the  SL are not taken into account and all the plasmons are intraband-like.

In the following we will consider two regimes. First, we fix the Fermi energy to $E_F$=0.1eV  ($E_F\gg E_G$) and then vary the superlattice potential $V_0$ from zero (uniform limit) up to $E_F$ where the electron density becomes zero at discrete locations, so-called singular points.\cite{Galiffi} By this, we analyze how the collective excitations change as the system becomes more and more inhomogeneous. Second, we analyze a system with average zero density ($E_F\ll E_G$) where the SL potential creates alternating  $p$- and $n$-doped regions. In each of these regimes,
we will analyze the dispersion of the first plasmonic sub-band and  
the plasmonic resonance that appears in the second sub-band  at the center of the Brillouin zone. This zero momentum plasmon can be probed in optical transmittance and reflectance measurements.

\begin{figure}[t]
\includegraphics[width=7.1cm]{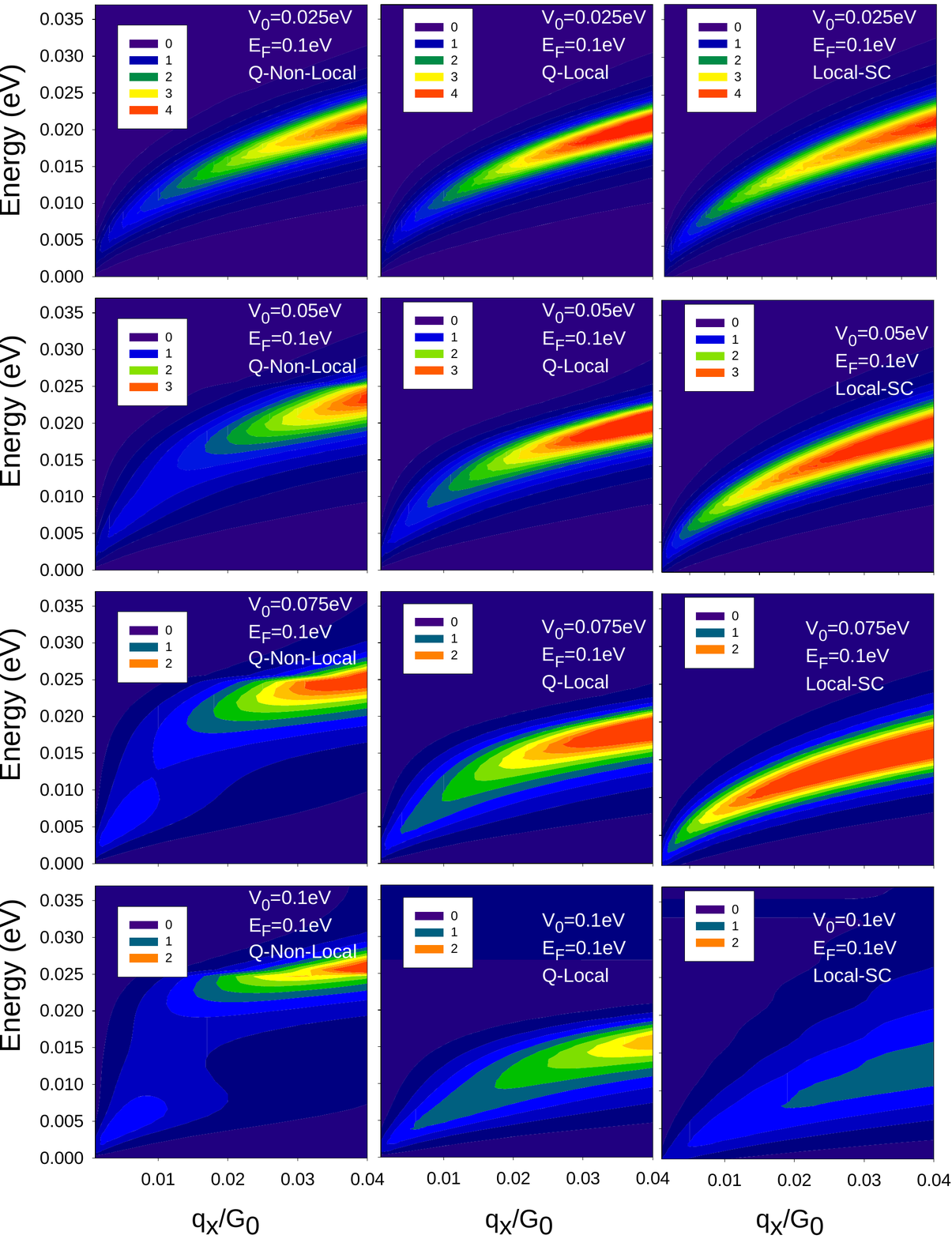}
\caption{Intraband plasmon dispersion(energy loss) with (left) and without (middle) non-local effects for a superlattice with period $L$=600$a$ and different values of $V_0$, as obtained in the quantum mechanical calculation.
The right panels show the
local semi-classical results. The comparison along  the $y$-direction is omitted since the Drude weight does not depend on $V_0$.}
\label{Figure2}
\end{figure} 
\vspace{0.2cm}
\par \noindent 
{\it Finite $E_F$: First plasmonic sub-band.} Let us first discuss the loss function for collective excitations with momentum $q_y$. We observe a well-defined low energy branch that disperses as   
$\sqrt { \frac {D_0 }{2\epsilon\epsilon _0}  q _y } $ with the same Drude weight as obtained from pristine graphene
. This part agrees with the SC approximation.

On the contrary, the Drude weight in the direction of the superlattice is reduced for increasing $V_0$ due to transfer of spectral weight from interband to intraband transitions.\cite{SI} This is also manifested in the loss function which is shown on the left column of Fig. \ref{Figure2} for various amplitudes $V_0/E_F$. Increasing the modulation $V_0$ renders the plasmons ill-defined below an energy threshold which we identify as $E_G$. 

A large dissipative plasmonic response at low energies is also seen in metallic structures with nanogap features.\cite{Yan15,Zhu16} However, the Q-local and semi-classical Thomas-Fermi approximations fail to predict this threshold energy $E_G$, as shown in the centre and right panels of Fig.\ref{Figure2}.

The vanishing of the loss function is the result of new decay channels that develop due to electronic band-folding that  allows  for electron-hole transitions at  energies $\hbar\omega\lesssim E_G$. An equivalent view point is that umklapp processes scatter the plasmonic excitations with $\q\sim0$ into the particle-hole continuum of intraband transitions where they decay due to strong Landau damping. This is only possible for $\hbar\omega\lesssim E_G$ and a purely non-local and quantum-mechanical effect since plasmonic band folding is only encoded in $\sigma_{G,G}^{\nu\nu}$ with $G\neq0$. It is thus even absent in the quantum-local (Q-local) approximation $\sigma_{G,G'}^{\nu\nu}\to\sigma_{G-G',0}^{\nu\nu}$. For energies $E \gtrsim E_G$, the loss function  becomes isotropic, Fig. \ref{Figure3}(a)-(b), and displays a well-defined Drude-like dispersion, with a Drude weight value equal to the intraband Drude weight in the $y$-direction.

Remarkably, the anisotropy in the plasmon dispersion is opposite to that occurring in the electronic band structure of the SL,\cite{SI,Park:2008aa,Barbier:2008aa,Arovas_2010,Burset_2011,Brey_2009}  where, due to Klein tunneling the velocity of the electrons moving along the SL axis is not modified, whereas it is strongly reduced in the perpendicular direction. The reason is that plasmons are collective excitations which even if characterised by a wave vector parallel to the superlattice are composed by electrons and holes moving in a range of directions. Plasmons are thus hardly affected by the Klein paradox.
 
\vspace{0.2cm}
\par
\noindent
{\it Finite $E_F$: Second plasmonic sub-band.}
In Fig.\ref{Figure3}(a)-(b), we plot the energy loss function at the center of the Brillouin zone for a SL with $V_0=E_F/2$. 
In fact, the loss function presents a double peak structure, see inset of Fig.  \ref{Figure3}(a), which widens and becomes less intense for increasing $V_0$.
The double peak structure reflects the folding onto ${\bf q}$=0 of two
states with momentum $\pm G_0$. These states interact between them presenting a small energy splitting proportional to $V_0$, as can be seen in inset of Fig. 2 and also in the SM.\cite{SI} 
\begin{figure}[t]
\includegraphics[width=8.cm,clip]{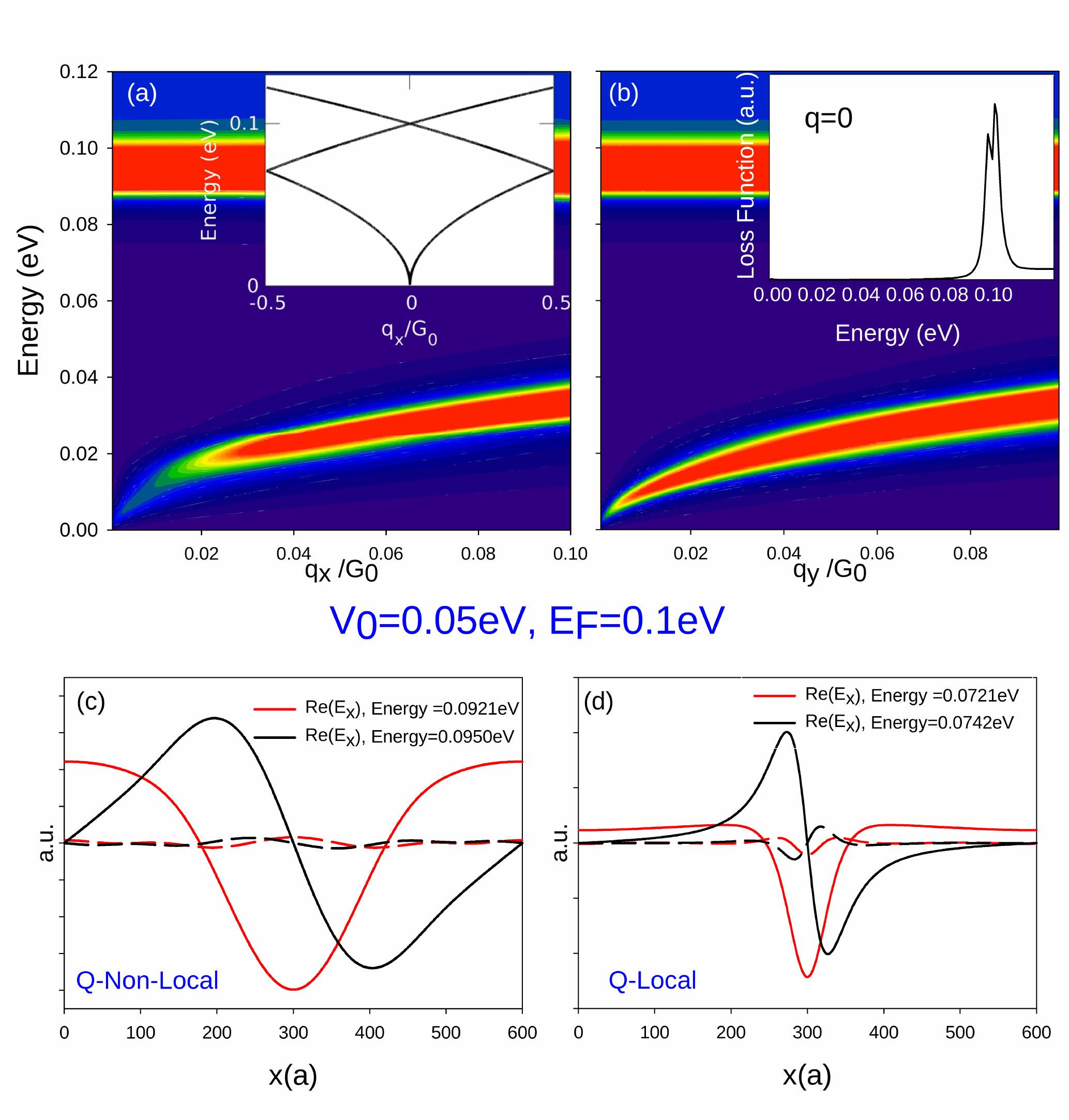}
\caption{Quantum non-local energy loss as function of (a) $q_x$ and (b) $q_y$ for a SL of period  $L=600a$, Fermi energy $E_F=0.1$eV and  amplitude $V_0$=0.05eV. {In the inset of panel (a), we plot the idealized illustration of the folding mechanism for the emergence of the second plasmonic subband.} In the inset of panel (b), we plot the energy loss for ${\bf q}$=0. Panel (c) shows the real (full lines) and imaginary (dashed lines) part of the electric fields, respectively, of the two strongest sub-band dielectric eigenmodes at ${\bf q}$=0.
In panel  (d), we plot the same as in panel (c), but for the eigenmodes obtained in the Q-local approximation. } 
\label{Figure3}
\end{figure} 

The modes in the second plasmon sub-band have a strong 
contribution from electronic 
interband transitions generated by the SL potential. This contribution increases as the SL perturbation $V_0$ increases.
 Nevertheless, the energy of the second plasmonic sub-band is very close to 
 $\hbar \omega _p (G_0)$, i.e. related to the Drude weight evaluated at the Fermi energy. The energy difference with respect to Eq. (\ref{Plasmon}) can be explained by non-local effects of the conductivity,\cite{Wunsch:2006aa,Hwang:2007aa} see SM.\cite{SI}
{
In Fig. \ref{Figure3}(c), we plot the $\hat x$-component of the electric field corresponding to the two eigenmodes that appear in the inset of Fig. \ref{Figure3}(a). The electric fields have the form of sine and cosine functions and thus correspond to the combination of the folded states with momentum $\pm G_0$, showing that the plasmon modes can be obtained from folding of the unperturbed plasmons, even in this case of large amplitude of the perturbation. The length scale of the interband plasmons is given by the period of the superlattice, i.e., $L\sim1/G_0$, and are thus not confined.
In Fig. \ref{Figure3}(d), we plot the electric fields corresponding to the double peaks that
appear in the Q-local approximation. Again, the two dielectric modes correspond to even and odd functions, but  the electric fields are localized in the region of low density as the SC local approximation also predicts.\cite{Galiffi}

Hence, non-local effects allow plasmons to explore all spatial regions and become extended over the whole system, making the SL behave as a meta-material displaying a homogeneous optical response even in presence of singular points where the electron density vanishes. This contrasts with the results obtained in local approximations (both SC and quantum) which are especially drastic near singular points.

\vspace{0.1cm}
\par
\noindent
{\it $E_F$=0: Neutral  plasmons.}
For neutral graphene, we have $D_0=0$ at zero temperature, signaling the lack of plasmons. However, at finite temperature there is an effective Drude weight and plasmons can be defined.\cite{Vafek06,Gomez09,Stauber12} In this work, we show that in neutral, but modulated graphene, band-folding leads to the appearance of  plasmon sub-bands for $V_0\gtrsim E_G$.  Since for undoped samples, electron-hole symmetry with respect to the chemical potential implies that the system responds just to the absolute value of the SL potential, there is a halving of the periodicity of the modulation, shown to be an exact symmetry in the SM.\cite{SI} Therefore, the electronic sub-band energy scale is in this case related to $E_{G}\to2\hbar v_F G_0$. In the first plasmonic sub-band, the dispersion shows a strong anisotropy: as in the case of finite doping, we find a plasmonic gap  in $x$-direction, whereas unperturbed intraband plasmons in $y$-direction, see Fig. 4 in SM.\cite{SI}

In the center of the Brillouin zone, we find that for  large values of the modulation $V_0$, there exits a peak in the loss function that indicates the existence of well defined plasmonic resonances, see Fig. 8 of the SM.\cite{SI}   When decreasing $V_0$, the energy of the plasmon is red shifted, the peak broadens and a continuous background appears at high energies, eventually signalling the absence of plasmons. 
In Fig. \ref{Figure6}, we plot the energy of this resonance 
as function of the folded momentum 2$G_0$ in units of the
effective Fermi energy  ${\overline E}_F$=$\frac 2 {\pi} V_0$ and  effective Fermi wavevector
${\overline k}_F$=$\frac {{\overline E}_F }{\hbar v_F}$, respectively. The red dots thus correspond to peak positions for different values of $V_0$ and follow the Drude dispersion of pristine graphene of Eq.(\ref{Plasmon}) evaluated at ${\overline E}_F$ (blue dashed line).

The same behaviour can be seen for different periodicities $L=900a$ (magenta dots) and $L=300a$ (green dots) which demonstrates that electronic interband transitions in fact mimic an effective classical Drude weight $D\sim V_0$.
However, as shown in the inset of Fig. \ref{Figure6}, the effective classical Drude weight is red-shifted by quantum effects associated to interband transitions and are related to $E_{G}$. For a  more detailed discussion, see SM.\cite{SI}.

We finally note that one-dimensional edge plasmons have been predicted to exist at the interface of a $p-n$ junction which show an unusual $q^{1/4}$-dispersion.\cite{Mishchenko10} We do not observe this effect, because their existence requires wavelengths much larger than the width of the junction, while here we have addressed the opposite limit. 

\begin{figure}[t]
\includegraphics[width=7.5cm]{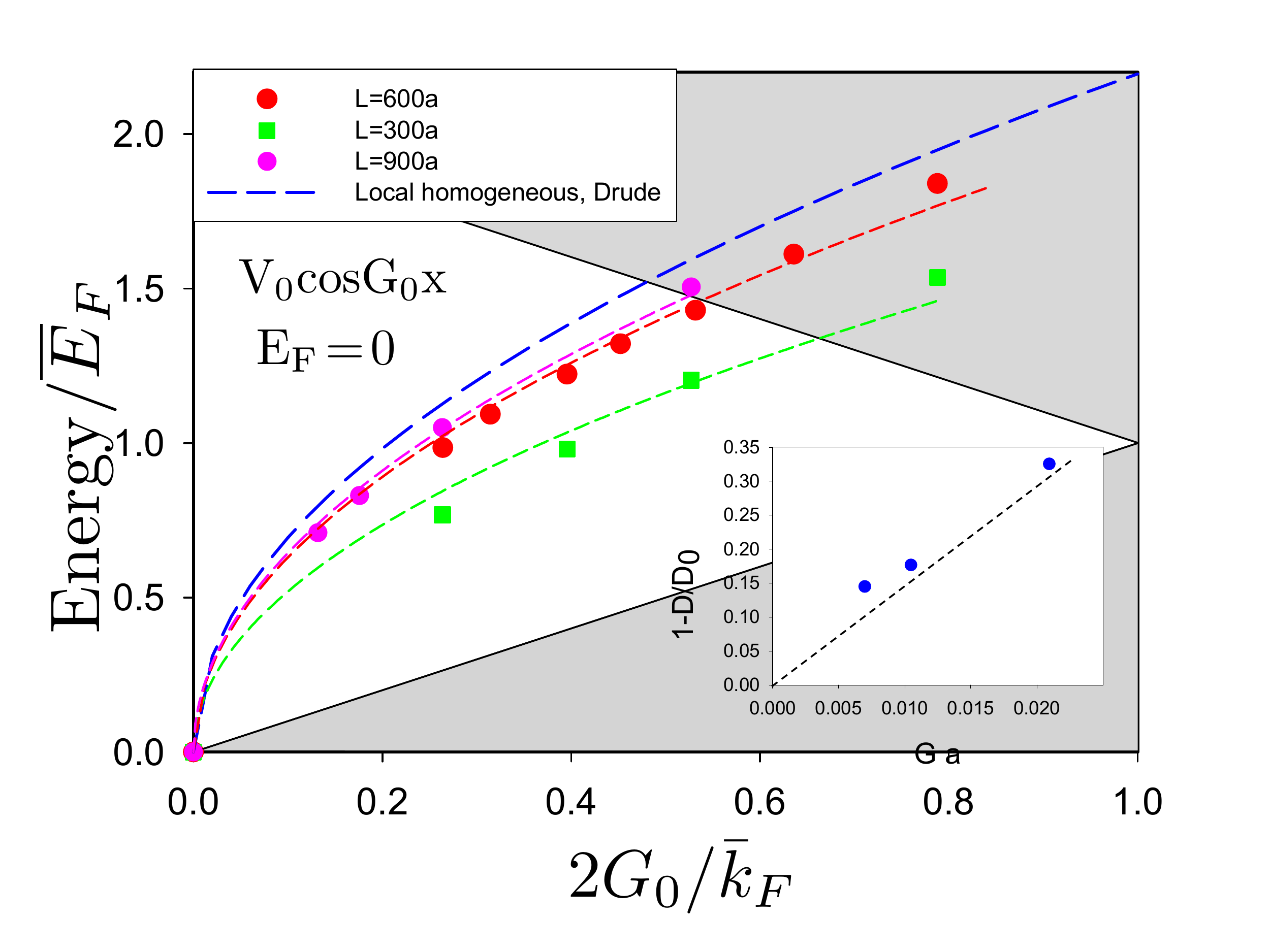}
\caption{Second sub-band plasmonic energies for SL potentials with period $L$=900$a$ (magenta), $L=600a$ (red) and $L=300a$ (green) for different values of $V_0\leq0.2$eV. The grey regions indicate the existence of electron hole damping by intra or interband (electronic) excitations. The dashed lines are the corresponding fits to Eq. (1) extracting the effective Drude weight $D$. Inset: The effective Drude weight as function of the reciprocal lattice vector. 
} 
\label{Figure6}
\end{figure}

{\it Summary of different Drude regimes.} As noted above, the zero momentum finite energy  plasmonic excitations can be described by Eq. (\ref{Plasmon}), albeit with an effective wavenumber $G$ and effective Drude weight $D$. The actual values depend on the ratio between the Fermi energy $E_F$, the modulation $V_0$, and the quantum electronic  sub-band energy  $E_{G}=\hbar v_F G$. 
For $E_F > V_0$, the wavevector $G$ coincides with the SL reciprocal lattice vector $G_0$ whereas  
for $E_F\ll V_0$, a periodicity halving   occurs\cite{Wang09} and the relevant wavevector is $G=2G_0$. 

The different Drude regimes are illustrated in Fig. \ref{FigurePD} and defined in the following way:

(i) For $\hbar\omega< E_G$, the Drude weight in the $\nu$-direction is given by\cite{Stauber:2013aa}
\begin{equation}
D_{\nu}=\lim_{\omega \to 0}\omega {\rm Im} \sigma^{\nu\nu}_{G=0,G=0}(q=0,\omega)\;.
\label{Drude}
\end{equation}
In the limit $V_0\to E_F$, the Drude weight becomes highly anisotropic and the plasmons ill-defined in the direction parallel to the SL, $D_{x}\sim0$. 

(ii) For $\hbar\omega> E_G$, the isotropic Drude weight is given by
\begin{align}
D=\frac {e^2}{\hbar^2}\frac{\overline{|E_F+V(x)|}}{\pi}\;,
\label{DrudeAverage}
\end{align}
where average over one unit cell of the superlattice is implied. Moreover, the plasmon dispersion is isotropic, $D\sim D_{x}\sim D_{y}$ and the Drude weight 
agrees well with 
Eq. (\ref{Drude}) in $\hat y$-direction.
For $E_F\ll E_{G}$, a red-shift $\sim E_{G}$ needs to be included in Eq. (\ref{DrudeAverage}), see inset of Fig. \ref{Figure6}.

\begin{figure}[t]
\includegraphics[width=6.7cm]{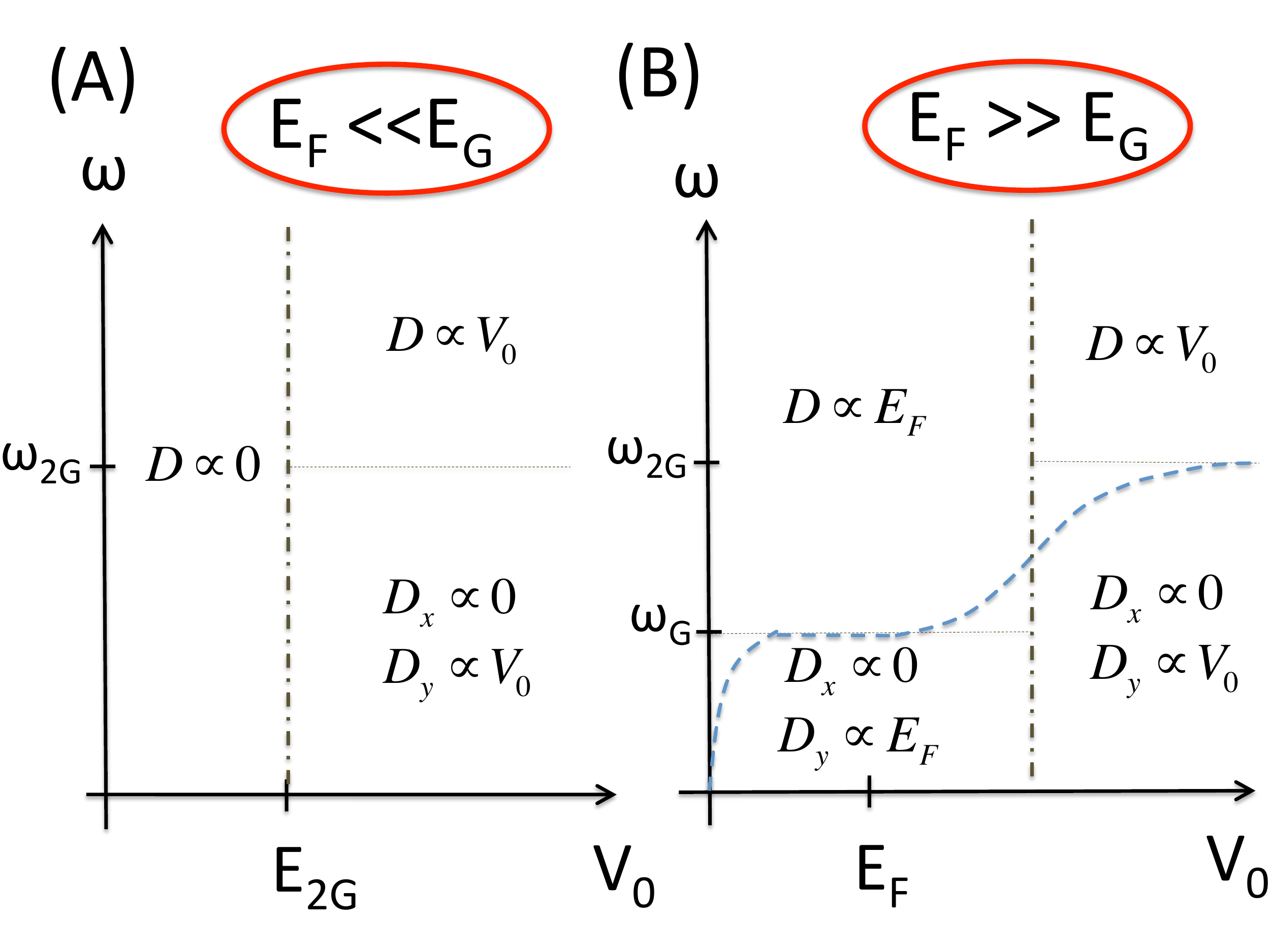}
\caption{Schematic map of the Drude weights for $E_F\ll E_G$ and $E_F\gg E_G$. All lines indicate crossover behaviour. We assume electron doping with $E_F>0$ and non-local or lattice corrections are not included.}
\label{FigurePD}
\end{figure}

{\it Conclusions.} We have analyzed the plasmonic properties of graphene dressed by a one-dimensional  superlattice potential. Even though the superlattice potential leads to the emergence of multiple electronic sub-bands, the character of the plasmonic excitations is not changed, i.e., they remain Drude-like charge-density oscillations where only the charge stiffness enters as effective parameter. In contrast to what both SC and Q-local approximations predict,
only minor corrections related to $V_0$ are observed when the modulation potential reaches the singular-point regime, $V_0 \sim E_F$.
In the case of  a neutral SL formed by periodically alternating $n$- and $p$-doped regions, we observe a halving of the periodicity, indicating that plasmons are insensitive to the sign of its carries, in agreement with exact results.\cite{SI} More interestingly, we again find Drude-like behaviour of the sub-band resonances related to $\omega_p(2G_0)$. 

We find that quantum electronic inter sub-band transitions  damp the propagation of plasmon excitations along the  SL axis, questioning the use of semiclassical calculations for 
this low frequency regime. {The classical approach is valid whenever the external grating does not induce a sufficiently large modulation of the "local" Fermi energy. In a quantum mechanical treatment, the Fermi energy is constant over the whole sample and the electronic density becomes inhomogeneous instead. This variation can be controlled by bringing e.g. the metal gratings closer to the graphene layer in order to enhance the screening effects and thus the local depletion of the electronic density. A rough estimate would be given by $V_0/E_F\sim0.1$. Another scale to contrast the classical approach is given by $E_G$ which is entirely due to band-folding. Thus, reducing the periodicity will eventually lead to deviations and a rough estimate is given by $L \sim1\mu$m.} In the case of plasmons with wavelengths of the order of the SL period, non-local effects reflected by the wave-nature of the electrons delocalize the plasmons in the SL unit cell, making it possible to describe the system as meta-material. 

Finally, let us point out that the collective excitations show an opposite behaviour to that of single-particle Dirac excitations, i.e., plasmons are largely unaffected by the superlattice in direction perpendicular to the modulation, but strongly modified in parallel direction. Klein tunnelling thus has no or very little effect on the plasmon propagation, contrary to several claims in the literature.

{\it Acknowledgments.} This work has been supported by Spain's MINECO under Grant No. PGC2018-097018-B-100, PGC2018-096955-B-C42, FIS2017-82260-P, MAT2017-88358-C3-1-R, and 
MDN-2014-0377 as well as by the CSIC Research Platform on Quantum Technologies PTI-001. LMM acknowledges Arag\'on Government through project Q-MAD.
\\
\\
\newpage
\begin{widetext}
{\bf\huge Supplementary Information}\\
 
 \section{Hamiltonian}
We consider the period of the superlattice to be much larger than the
lattice parameter $a=2.46${\r{A}} of graphene and the amplitude of the perturbation $V_0$  much smaller than the energy bandwidth of the graphene $p_z$-orbitals. For these conditions and for a single valley and a given spin orientation,  the low energy electronic properties are well described by the massless Dirac Hamiltonian in the presence of an external potential,
\begin{equation}
H =\hbar v_F \left (-i \sigma _x \partial _x-i \sigma _y \partial _y \right )+ V(x) {\cal I}\;,
\label{hamiltonian}
\end{equation}
where $\sigma _x$ and $\sigma _y$ are the Pauli matrices, $v_F$ is the Fermi velocity, $\cal I$ is the identity matrix, and $V(x)$=$V_0 \cos G_0x$ is the SL potential. $G_0=2\pi/L$ is the reciprocal lattice vector and we mainly choose a periodicity of $L=600a$ in this work. However, to consolidate our results, we will sometimes also consider periodicities of $L=300a$ and $L=900a$.

\begin{figure}[htbp]
\includegraphics[width=10cm,clip]{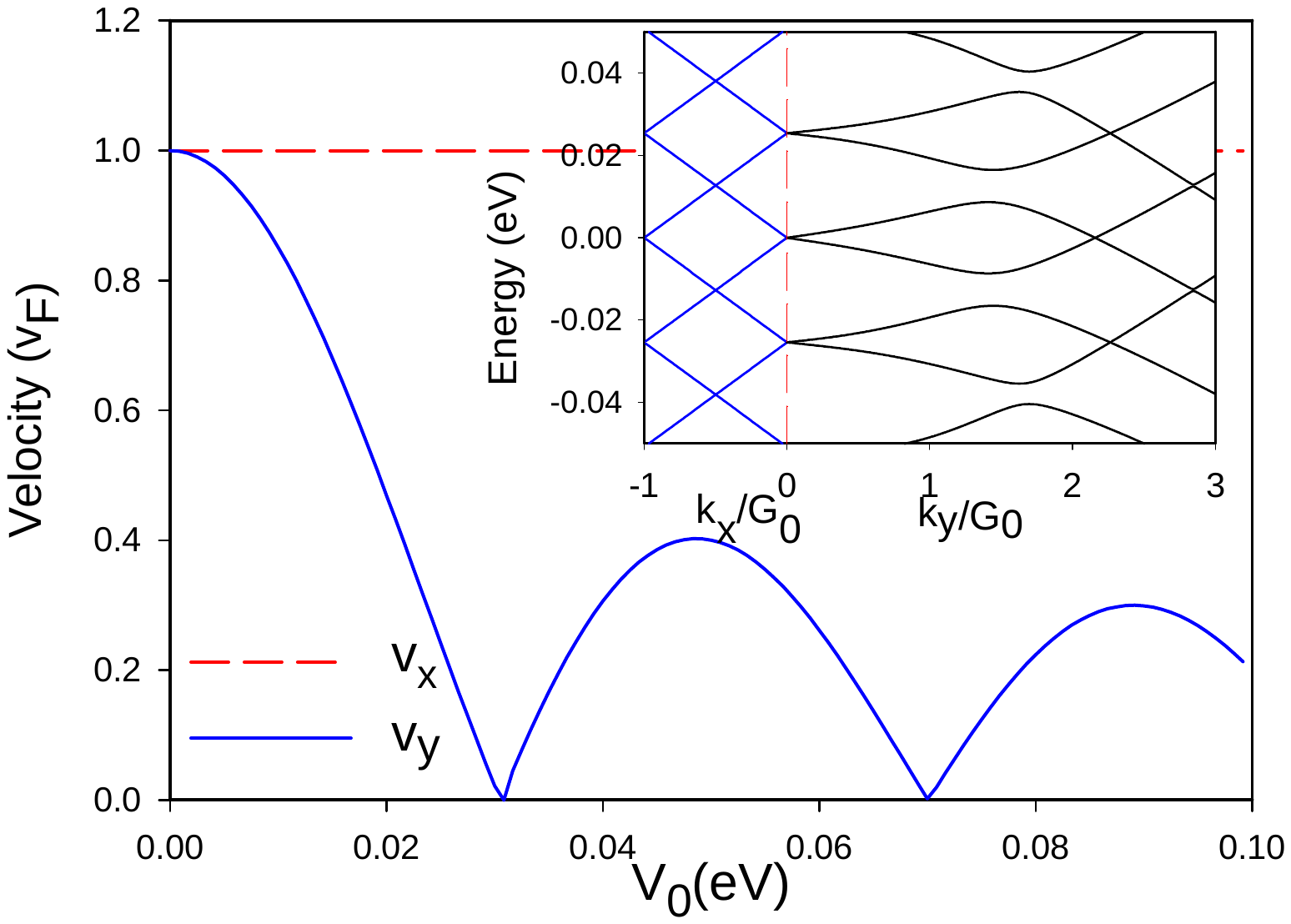}
\caption{Electronic velocity renormalization for a graphene superlattice with period $L$=$600a$ as function of the amplitude of the SL potential $V_0$. The velocity in the direction of the SL axis is not affected, whereas in the perpendicular direction,  it is extremely reduced, reaching sometimes zero. Inset:  electronic band structure for a superlattice with parameters $L$=600$a$ and $V_0$=0.1eV. The axis of the SL is along the $\hat x$-direction.} 
\label{Figure1}
\end{figure}  

The Hamiltonian acts on spinors whose components represent the amplitude on the two triangular lattices that form the graphene honeycomb lattice. The eigenstates of the system are characterized by a band index and the momentum ${\bf k}=(k_x,k_y)$ with $k_x$ restricted to the first Brillouin zone. 

Because of their chiral nature, the carriers in this system show an unexpected high anisotropic behavior,  see Fig. \ref{Figure1}. At the  Dirac point, the group velocity is extremely reduced in the direction perpendicular to the superlattice direction, whereas it is unchanged along the superlattice \cite{Park:2008aa,Barbier:2008aa}.
This anisotropy translates also in a high anisotropy of the electronic  transport\cite{Arovas_2010,Burset_2011}. Let us emphasise that superlattice potentials can be used for manipulating graphene's band structure. In particular, they can be designed to manipulate the number of zero energy modes  \cite{Brey_2009,Park_2009}.

We will find that the renormalization of the velocity of the electrons in the direction perpendicular to the SL axis does not affect the dispersion of the intraband plasmons in that direction, i.e., plasmons with $q_y$ will follow a typical Drude-like dispersion relation. The opposite occurs in the direction parallel to the SL axis, i.e., the plasmons with $q_x$ at low energies are slowed down and strongly damped even though the velocity of the electrons is not modified by the perturbation. 

We conclude that the Klein paradox does hardly affect the motion of the plasmon, mainly because plasmons are collective charge oscillations composed of multiple particle-hole pairs $c_{\k+\q}^\dagger c_{\k}$. The chirality of the electronic wave function is thus averaged out by summing over all directions of $\k$ similar to pristine graphene and two-dimensional topological surface states in Bi$_2$Se$_3$. For a more explicit discussion, see Ref. \cite{Stauber17}.

{The new damping mechanism acting on the low-energetic plasmons can be understood from the electronic structure. Due to the band-folding of the electronic dispersion, new interband transitions arise as can be seen on the left panel of Fig. \ref{Figure1a}. The transition matrix elements are strongest for $\hbar\omega\lesssim E_G$ with $E_G=\hbar v_FG_0$, the energy related to the first umklapp process.}

{This damping mechanism can also be understood from the two-particle spectrum of unperturbed graphene. For first-order umklapp-processes, energy is conserved, but the momentum changes by the reciprocal lattice vector $G_0$. These processes will thus scatter the plasmonic modes for $\hbar\omega\lesssim E_G$ into the electron-hole continuum, leading to the observed strong damping and schematically shown on the right panel of Fig. \ref{Figure1a}.}

\begin{figure}[htbp]
\includegraphics[width=10cm,clip]{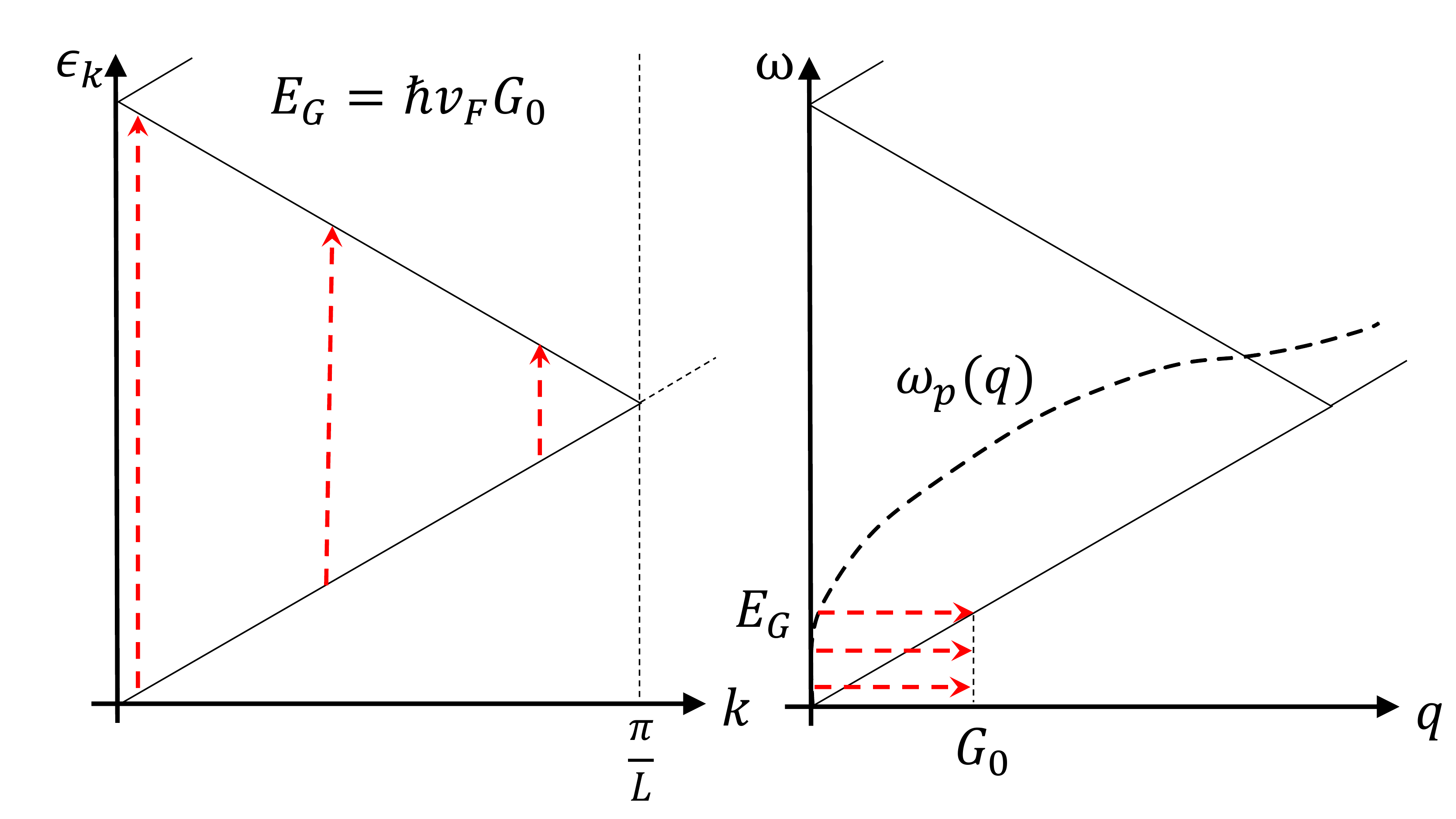}
\caption{Left panel: Schematic single-particle electronic spectrum in $x$-direction and the additional interbank transitions due to band-folding. Right panel: Schematic two-particle electronic spectrum indicating the scattering of the plasmonic mode into the electron-hole continuum.} 
\label{Figure1a}
\end{figure} 

\section{Drude plasmons}
Plasmons are collective density oscillations present in almost all electronic systems. They are most easily described by a hydrodynamic approach based on the continuity equation and linear response
\begin{equation}
-i\omega\rho=-\nabla\cdot\j=-\sigma(\omega)\nabla\cdot\E=\sigma(\omega)\nabla^2\phi\;.
\end{equation}
Note that we assumed that $\sigma(\omega)$ was isotropic and did not have any spatial dependence. With the Fourier transform of the electrostatic potential, $\phi(\rr)=\phi_\q e^{\i \q\cdot\rr}$, and the Poisson equation $\phi_\q=v_q\rho_\q$ with the Coulomb propagator $v_q$, we can write the above equation as
\begin{equation}
(-i\omega-\sigma(\omega) v_\q q^2)\rho_\q=0\;.
\end{equation}
In the absence of dissipation, collective density oscillations are thus defined by the dispersion relation $\omega_p=$Im$\sigma(\omega) v_qq^2$ which holds for all dimensions. This approximation is sometimes labeled as local homogeneous.

The imaginary part is usually well approximated by Im$\sigma(\omega)=\frac{D}{\omega}$ where $D$ is the Drude weight. The Drude weight can be obtained of general isotropic systems with electronic dispersion relation $E(\k)\sim|\k|^\nu$ on the same footing. For low temperature, this yields\cite{Stauber:2014aa}
\begin{equation}
\label{localCurrentResponse}
D=\frac{e^2}{\hbar^2}\frac{g_sg_v\nu}{2}\frac{E_F}{2\pi}\;,
\end{equation}
with $g_s$, $g_v$, the spin- and  valley degeneracies, respectively. Note that this gives the same Drude weight for 2D parabolic bands ($g_v=1$ and $\nu=2$) and graphene ($g_v=2$ and $\nu=1$). 

With the 2D Coulomb interaction $v_q=\frac{e^2}{2\varepsilon_0\epsilon q}$ and $\epsilon$ the relative dielectric constant, the plasmon dispersion for a general 2D system in the local approximation is thus given by
\begin{equation}
\label{TwoDPlasmons}
\omega_p^D=\sqrt{\frac{D}{2\varepsilon_0\epsilon}q}\;.
\end{equation}
with the characteristic square-root dispersion. For graphene on the interface of two different dielectric media, one further has $\epsilon=(\epsilon_1+\epsilon_2)/2$.

\section{Optical conductivity and loss function in a periodic system}
In a periodic system, the response to an external momentum ${\bf q}$ involves all reciprocal lattice vectors ${\bf q}+G$. Within the mean field approximation, the longitudinal dielectric matrix has the form\cite{Vignale-book}
\begin{align}
{\epsilon}  ({\bf q}+G,{\bf q}+G'; \omega)&= \delta _{ G, G'} - v(|{\bf q}+{ G}|)  \frac 1 {i \omega e^2}\label{epsilon}\left [ { (q_x+ G) (q_x+ G')     \sigma ^{xx} ({\bf q}+G,{\bf q}+G';\omega)+
q_y ^2 \sigma ^{yy} ({\bf q}+G,{\bf q}+G';\omega)
}\right ]
\end{align}
where $G = n  G_0$, being $n$ an integer,  $v(q)$=$\frac {e ^2}{2\epsilon_0\epsilon q}$ is the Coulomb interaction in two-dimensions
and $\sigma ^{\nu\nu} ({\bf q}+G,{\bf q}+G';\omega)$ is the dynamical optical conductivity  of the system in the $\nu$-direction.

\subsection{Optical conductivity}
For non-interacting particles, the optical conductivity  is obtained from the Kubo formula and one gets the expression
\begin{equation}
{\sigma}^{\nu \nu}  ({\bf q}+G, {\bf q}+ G';  \omega)=-i \frac {\hbar}{S} \sum _{i,j} \frac {n_F(E_i)-n_F(E_j) }{E _j -E_i}
\frac {<i| {\hat v}_{\nu} e ^{-i( {\bf q}+G)\hat{\bf r}}|j><j|\hat v_{\nu} e ^{i ({\bf q}+G')\hat{\bf r}}|i>}  {\hbar (\omega +i \eta)+E_i -E_j}\;.
\label{sigma}
\end{equation}
In this expression, $|i>$ and $E _i$ are the eigenvectors and eigenvalues of the one-electron Hamiltonian Eq. (\ref{hamiltonian}), respectively, $n_F$ is the Fermi occupation function, and $\hbar \eta$ represents the quasiparticle lifetime broadening. The index $i$ represents the quantum number of the eigenstate including momentum, band index, and spin. Finally,
$\hat v_{\nu}$=$v_F \sigma _{\nu}$ is the velocity operator in the direction $\nu$ associated to the Hamiltonian of Eq. (\ref{hamiltonian}). 

The  nonzero
off-diagonal elements in Eq. (\ref{sigma})
are due to the spatial modulations of the carriers produced by the superlattice periodic potential, whereas the finite diagonal terms with $G \neq 0$ correspond to non-local corrections. 
The superlattice potential makes the system anisotropic inducing    $\sigma ^{xx} \neq \sigma ^{yy}$.  

For our numerical calculations, we need to limit ourselves to a maximal reciprocal lattice vector and we mainly choose $G_{max}/G_0$=30. We have checked that this plane wave cutoff provides accurate convergency by comparing our results with $G_{max}/G_0$=40. We further set $\eta=0.002$eV.

\subsection{Loss function}
The collective charge density  excitations (plasmons) are given by the zeros of the real part of the dielectric constant, Eq. (\ref{epsilon}). For a given wave-vector and frequency, the eigenstates 
$\phi _n({\bf r})$  associated to the eigenvalues  $\epsilon  _n( {\bf q},\omega)$ represent all  different dielectric eigenmodes of the system
that are orthogonal and electrodynamically decoupled.\cite{Baldereschi-1978,Car:1981aa}
The eigenmodes, $\phi_n$, have in general an imaginary part that indicates the changes in the phase of the plasmonic electrical potential due to the spatial variation of the optical conductivity\cite{Andersen:2012aa}. Also, the eigenvalues $\epsilon_n({\bf q},\omega)$ are complex numbers, and in this case the  imaginary part  represents the collective excitation broadening.  
If the imaginary part of the dielectric eigenmodes does not vary too much near the charge density excitation, the condition for the existence of a plasmon  is more accurately  defined 
by the condition that the quantity
$-{\rm Im} \frac 1{\epsilon_n({\bf q},\omega)}$ shows a local maximum.

The number of eigenvalues of Eq. (\ref{epsilon}) is given by the number of plane waves included in the calculation. From all the dielectric modes, we choose the eigenvalue $n_0$ with the largest values of $-{\rm Im} \frac 1{\epsilon_n({\bf q},\omega)}$\cite{Andersen:2012aa,Wang:2015aa}. 
For a given wavevector ${\bf q}$, frequency $\omega$, and $n_0\equiv n_0({\bf q},\omega)$\color{black}, we thus define the loss function as
\begin{equation}
S({\bf q},\omega)=-{\rm Im} \frac 1{\epsilon_{n_0}({\bf q},\omega)} \, \, .
\end{equation}

We have checked that the second largest values of this quantity are considerable smaller and we do not consider them.
The electric field corresponding to the eigenvalues $\varepsilon _n$ that defines the loss function does not change significantly  when the frequency is varied near the
position of the plasmon peak.

\section{Drude weight in modulated systems}
The optical conductivity contains both superlattice intra- and interband contributions and can be  written as
\begin{equation}
{\sigma}^{\nu\nu} ({\bf q }+G,{\bf q}+G' ; \omega)=i \frac {D_{\nu}}{\omega +i \eta}\delta _{ G, G'} \delta _{G,0}-
{\bbsigma}^{\nu\nu} ({\bf q }+G,{\bf q}+G' ; \omega)\;,
\label{RealSigma}
\end{equation}
where in $\bbsigma$ the intraband contributions are omitted. $D_{\nu}$ is the Drude weight in the $\nu$-direction.
Notice that in the presence of a superlattice potential, the interband contribution  includes both the intercone electronic transitions and the
transitions  generated by the  new periodicity. 

\subsection{Drude weight in Dirac systems}
\begin{figure}[htbp]
\includegraphics[width=10cm]{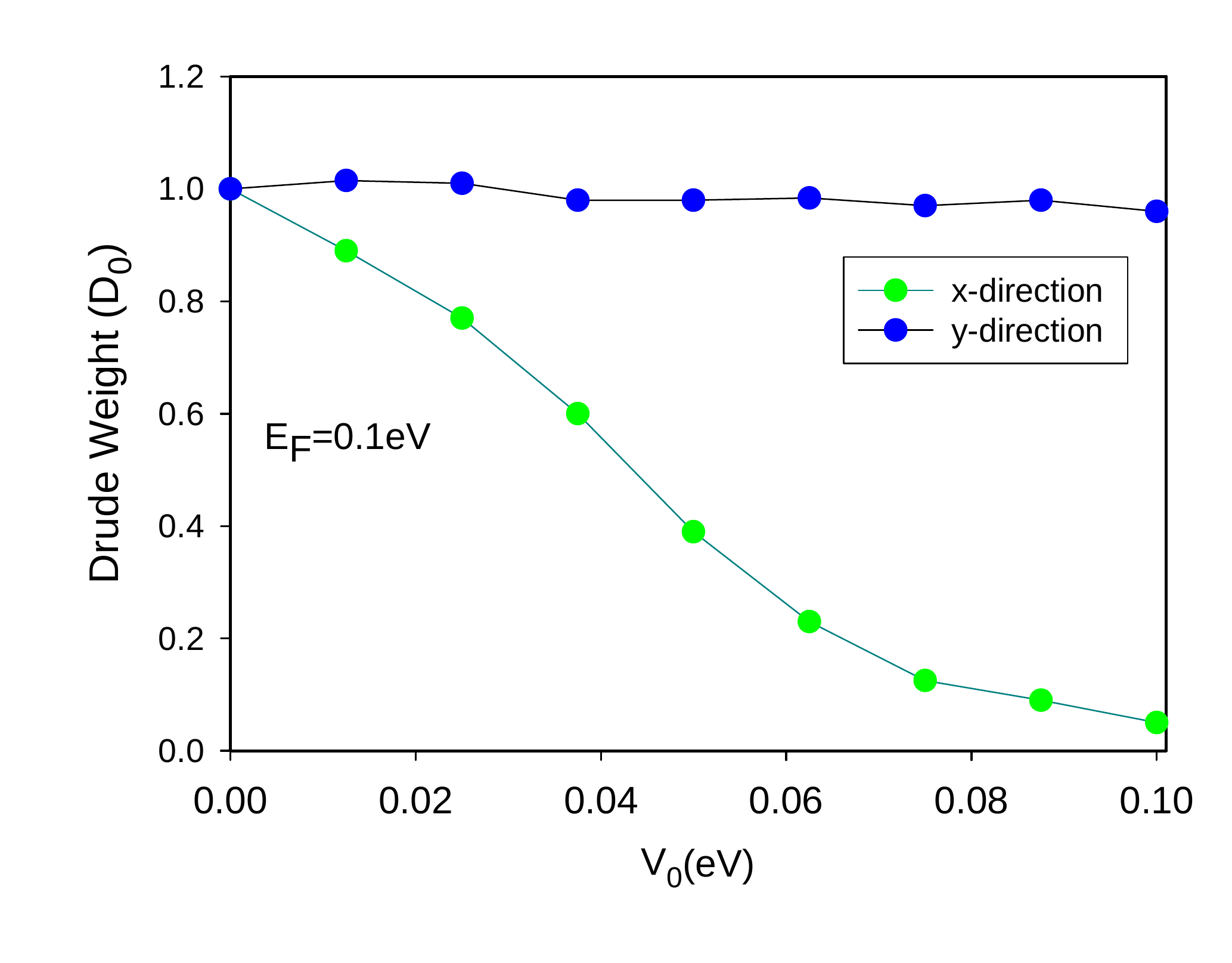}
\caption{Drude weight as function of the amplitude of the superlattice potential, $V_0$. 
$\hat x$ is the direction of the SL axis and $\hat y$ is the perpendicular direction. The Fermi energy is $E_F$=0.1eV and the period of the  superlattice is $L$=600$a$. Drude weights are given in units of $D_0$=$\frac {e^2}{\hbar^2} \frac {E_F}{\pi}$, i.e., the Drude weight of a uniform system with the same Fermi energy.}
\label{FigureDrude}
\end{figure} 

In order to calculate the complex optical conductivity, the simplest way is often to first compute its real part and then to
obtain the imaginary part by making use of the  Kramers Kronig relations.
In Dirac Hamiltonians, the bands are  linear in the momentum and extend up to infinity.  This, together with the chiral character of the
eigenstates, leads to a finite value of the real part of the conductivity at large frequencies.
For undoped  homogeneous graphene, the real part of the optical conductivity reads
\begin{equation}
\sigma_0 ({\bf q},{\bf q},\omega)=\frac {1}{4} \frac {e^2}{\hbar} \frac {\omega}{\sqrt{\omega^2-v_F^2 q^2}} \Theta(\omega-v_F q) \,\,.
\end{equation}
The contribution at high frequencies leads to a divergence of the imaginary part of the optical conductivity if computed through  the  Kramers-Kronig relation. We avoid this divergence following the procedure of ultraviolet field theories and define the imaginary part relative to this divergent contribution,\cite{Sabio:2008aa,Stauber:2013aa}
\begin{equation}
{\rm Im} {\sigma}^{\nu\nu} ({\bf q}+G,{\bf q}+G'; \omega)=\frac 2 {\pi \omega} {\rm P}\int _0 ^{\infty} d u \, {u}^2 
\frac {{\rm Re}\,{\bbsigma}^{\nu\nu} ({\bf q}+G,{\bf q}+G'; u)-\sigma_0(G,G;u)\delta_{G,G'}}{\omega ^2 -u^2} \, \, . 
\end{equation}
With this procedure, we obtain the following expression for the Drude weight:
\begin{equation}
D_{\nu}=\frac 2 {\pi} \int _0 ^{\infty} \!\! \!\! d \omega \left ({\rm Re} \sigma ^{\nu\nu} (0,0;\omega)- \sigma_0 \right)\;,
\label{Drude}
\end{equation}
where $\sigma_0=\frac{1}{4}\frac{e^2}{\hbar}$ denotes the universal conductivity of neutral graphene in the local limit. We note that this expression is numerically more stable than the equivalent definition 
\begin{equation}
D_{\nu}=\lim_{\omega\to0}\omega\sigma^{\nu\nu}(0,0;\omega)\;.
\end{equation}
The definition of Eq. (\ref{Drude}) assures that the  Drude weight or charge stiffness is zero for undoped, homogeneous graphene as it must. With $\Lambda$ denoting the (large) energy cutoff, this also yields the $f$-sum rule generally valid for Dirac systems,
\begin{equation}
\int_0^\Lambda \Re\sigma^{\nu\nu}(0,0;\omega)d\omega=\sigma_0\Lambda  \, \, . 
\label{sumrule}
\end{equation}
For isotropic Dirac systems, we have 
\begin{equation}
D_0=\frac{e^2}{\hbar^2}\frac{|E_F|}{\pi}\;.
\label{Plasmon}
\end{equation}
\begin{figure}[h]
\includegraphics[width=10cm]{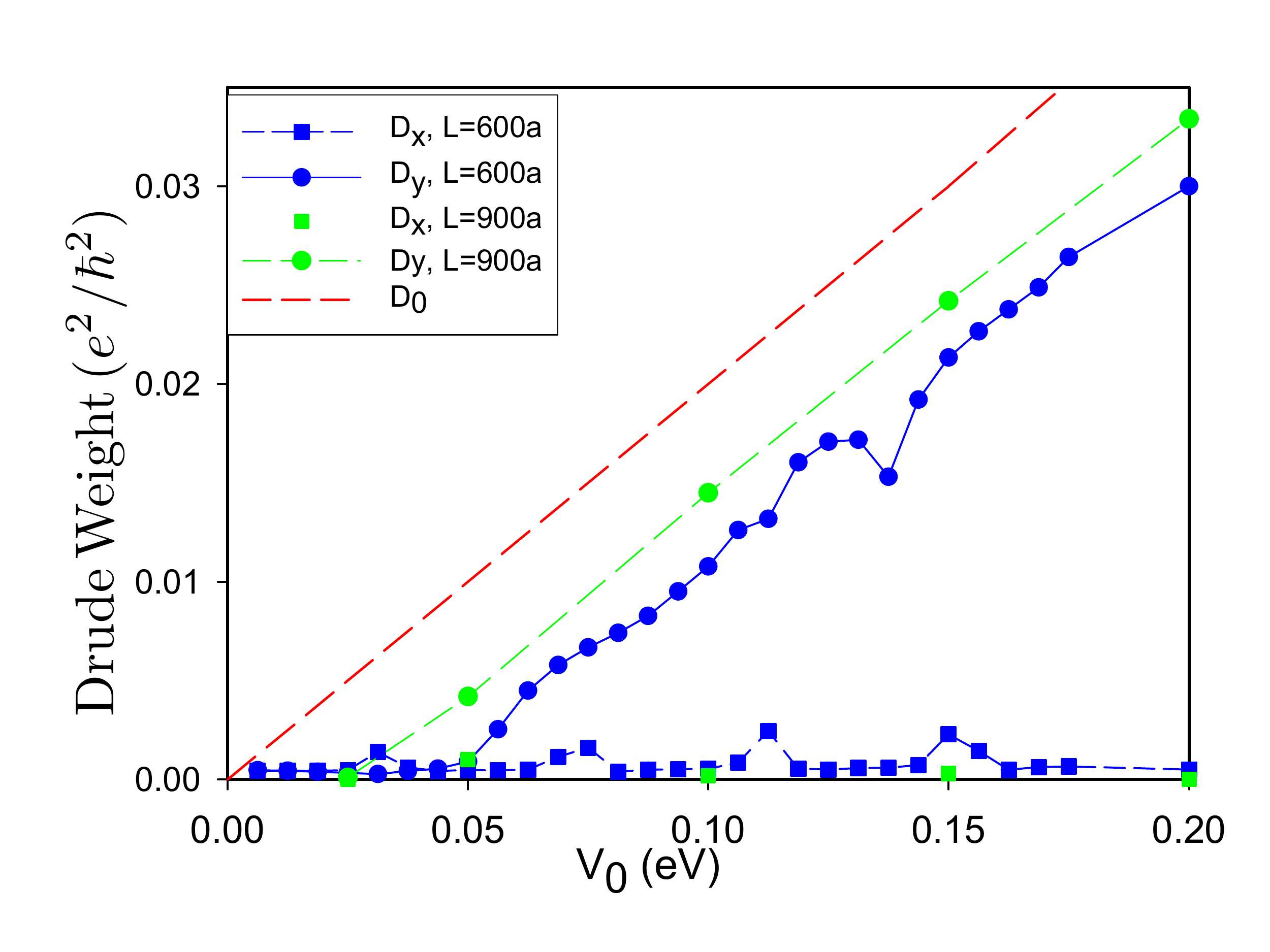}
\caption{Drude weight as function of the amplitude of the superlattice potential, $V_0$. 
$\hat x$ is the direction of the SL axis and $\hat y$ is the perpendicular direction. The Fermi energy is zero ($E_F$=0) and the period of the  superlattice is $L$=600$a$ (blue symbols). The Drude weight corresponding to ${\overline E}_F=\frac{2}{\pi}V_0$ of a homogeneous system, i.e., $D_0$=$\frac {e^2}{\hbar^2} \frac{{\overline E}_F}{\pi}$, is shown as dashed line. For comparison, also the Drude weight for $L=900a$ is shown for some SL potentials $V_0$ (green symbols). All Drude weights are in units of $\frac {e^2}{\hbar^2}$eV.}
\label{FigureDrudeN}
\end{figure} 
\begin{figure}[h]
\includegraphics[width=10cm]{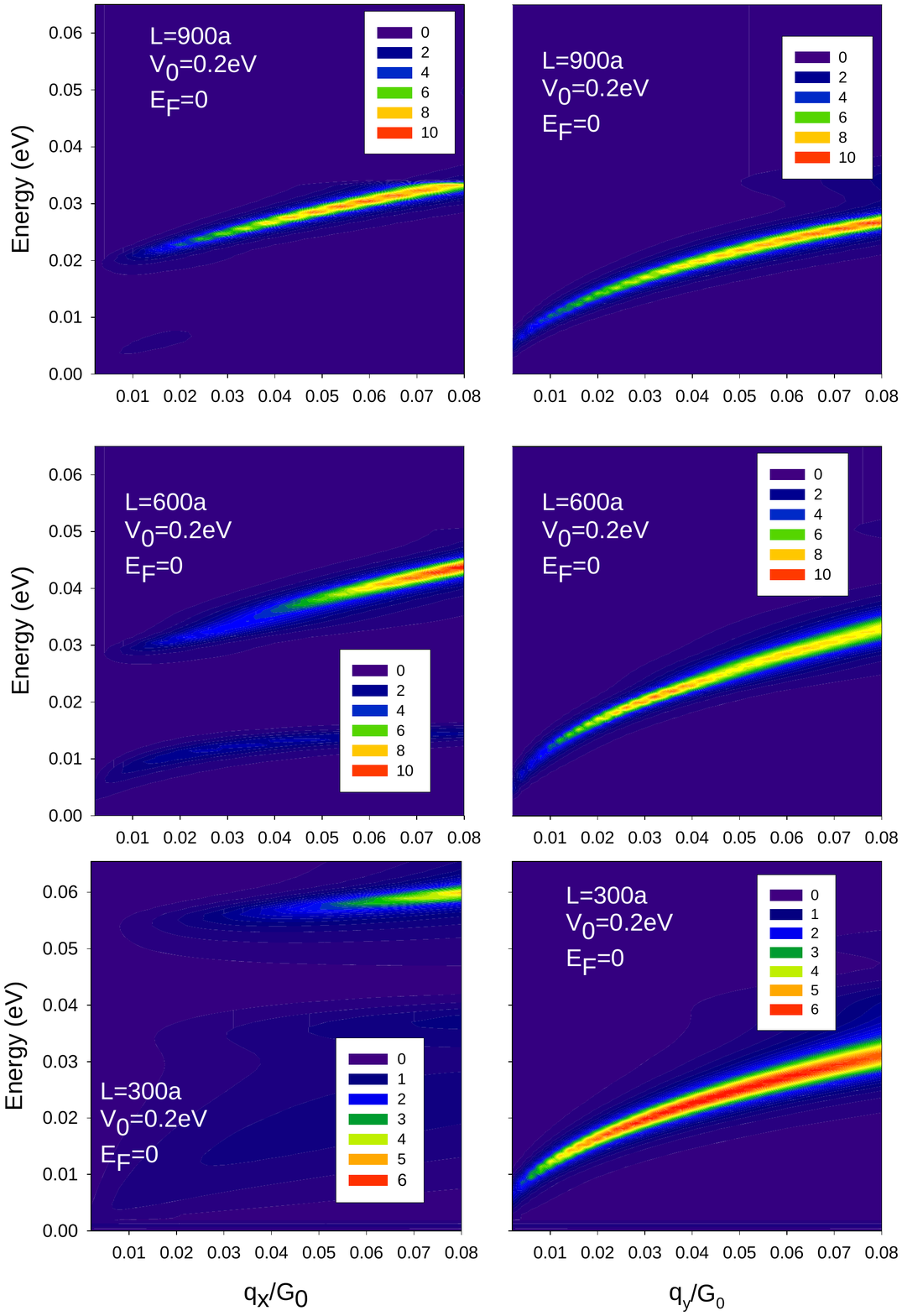}
\caption{
Energy loss function at $E_F=0$ and $V_0=0.2$eV for a SL potential with various periods $L/a$=300,600,900 in the $\hat x$-(left) and $\hat y$-direction (right). }
\label{FigureCutoff}
\end{figure} 

\subsection{Drude weight of a SL at finite charge density}
In Fig. \ref{FigureDrude}, we plot the Drude weights in the $\hat x$- and $\hat y$-direction as obtained from Eq. (\ref{Drude}) for  $E_F$=0.1eV as function of the SL potential, $V_0$. We plot the  Drude weight in units of  $D_0$=$\frac {e^2}{\hbar^2} \frac { E_F}{\pi}$ that corresponds to the Drude weight of a uniform system with the same Fermi energy, Eq. (\ref{Plasmon}). In the direction perpendicular to the SL axis, the Drude weight is practically unaffected by the perturbation, indicating  that in this direction the low energy intraband plasmons are not altered by the SL. 

On the contrary, the Drude weight in the direction of the superlattice is reduced  with respect to  the value of pristine graphene, $D_0$. This reduction increases with $V_0$ and has its origin in the transfer of spectral weight from low energy electronic interband transitions to higher energy electronic intraband transition allowed by the periodicity of the SL potential, also seen in twisted bilayer  graphene.\cite{Stauber:2013aa} The intensity of the low energy intraband plasmon along the SL axis is very weak and a strong reduction of spectral weight is seen in the loss function presented in Fig. 1 of the main text.

\subsection{Drude weight and loss function of a SL at zero charge density}
In a neutral system, the charge density is zero and in a homogeneous system the compressibility or charge stiffness must be zero as well. Still, in a neutral, but modulated system the Drude weight can become finite in the direction perpendicular to the SL potential. Interestingly, this happens only beyond a critical modulation $V^*=E_{2G}$, the energy scale that defines umklapp processes just as in the doped case. 

For $V_0>V^*$, the Drude weight is proportional to the effective Fermi energy ${\overline E}_F=\frac{2}{\pi}V_0$, see Fig. \ref{FigureDrudeN}:
\begin{equation}
D_{y}=\frac{e^2}{\hbar^2}\frac{{\overline E}_F-E_{2G}}{\pi}
\end{equation}

In the $\hat x$-direction, the Drude weight is zero indicating a gap in the plasmonic spectrum. We also observe spikes at certain values of $V_0$ which coincide with the zeros of the Fermi velocity as shown in Fig. \ref{Figure1}, where the Drude weight defined as inverse mass would formally diverge.

The discussion of the Drude weight at charge neutrality needs to be complemented by a discussion of the energy loss at low frequencies, i.e., below the energy of the second plasmonic sub-band. In Fig. \ref{FigureCutoff}, we show the energy loss function for SL potentials with various periods $L/a$=300,600,900 in the $\hat x$ (left) and $\hat y$ (right) direction. 

In the $\hat x$-direction, we observe a plasmonic gap that scales with $E_{2G}$ and we associate this to the opening of new emerging decay channels due to the superlattice. Another view point is that umklapp processes with energy $E_{2G}$ scatter the plasmonic excitations with $\q\sim0$ into the particle-hole continuum of intraband transitions. We also observe a sub-gap plasmonic band with less intensity, most clearly developed for $L=600a$. 

In the $\hat y$-direction, there is no gap in the plasmonic band and a decrease of the spectral weight can be observed when increasing $G$ (note the different colour scale in the case of $L=300a$). This is consistent with the energy loss of a dissiplationless Drude system reflecting the decrease of the Drude weight by $E_{2G}$:
\begin{equation}
S(q,\omega)=\frac{\pi}{2}\omega_p\delta(\omega-\omega_p)\;,
\end{equation}
with the plasmonic energy given in Eq. (\ref{TwoDPlasmons}).

\subsection{Drude weight of a SL at zero charge density and band structure}
On physical grounds, the Drude weight for our non-interacting Hamiltonian should coincide with the (inverse) mass tensor and, therefore,  could also be obtained directly from the band structure as
\begin{equation}\label{Dband}
D_{\nu}=  \frac{e^2}{S} \sum_{\bm k,n} |\mel{\bm k,n}{\hat{v}_{\nu}}{\bm k,n}|^2 \delta (E_{\bm k,n}-E_F)
,\end{equation}
where $\hat{v}_{\nu}=v_F \sigma_{\nu} $ are graphene velocity operators along $\nu=x,y$. 
The band structure of modulated neutral graphene, $E_F=0$, corresponds to a non metal: the Fermi surface shrinks to a finite number of Dirac points along the line $k_x=0$ in reciprocal space. Its number depends on the size of the modulation, $V_0$, and some of them are seen in the inset of Fig. \ref{Figure1}. Around some of these points, the bands can be  {\it extremely} anisotropic, with almost no dispersion along $x$, and numerical care is required to show that Dirac points remain points and do not become Fermi lines. 

The problem is that the zero density of states at $E_F=0$  implies  zero Drude weight from Eq. \ref{Dband}. Therefore, the finite  Drude weight obtained from our Kramers-Kr\"onig (KK) procedure for $E_F=0$, see Fig. \ref{FigureDrudeN}, presents us with a potentially severe problem, if taken prima facie.

It turns out that we can solve this apparent contradiction in a physically satisfactory way, further clarifying the meaning of our calculation and its limits. Our KK calculation includes a phenomenological damping, $\eta=.002 $eV, mostly for practical reasons. This means that we are killing all response structure for frequencies below this damping. Physically, this must be equivalent to saying that the KK calculation should start to resemble the response of the {\it clean} system for frequencies at least of the order of the damping, and at finite frequency the response is finite. Our  damping is {\it nominally} small but, as mentioned before, enough to render a finite value for the Drude weight within the KK formalism.

To show that the KK calculation of the Drude weight at zero charge density amounts to introducing a finite energy scale, we can turn the argument around, and inquire about the fate of the Drude weight as calculated from the band structure, Eq. \ref{Dband}, when a finite energy scale is introduced. On physical grounds, this can be done by switching the temperature on, for instance. This amounts to performing the following replacement in Eq. \ref{Dband}:   
\begin{equation}\label{T}
\delta (E_{\bm k,n}-E_F) \to \frac{1}{4 kT \cosh^2(\tfrac{E_{\bm k,n}-E_F}{2 kT})}
.\end{equation}
%
\begin{figure}[htbp]
\begin{subfigure}{.45\textwidth}
\includegraphics[width=8cm]{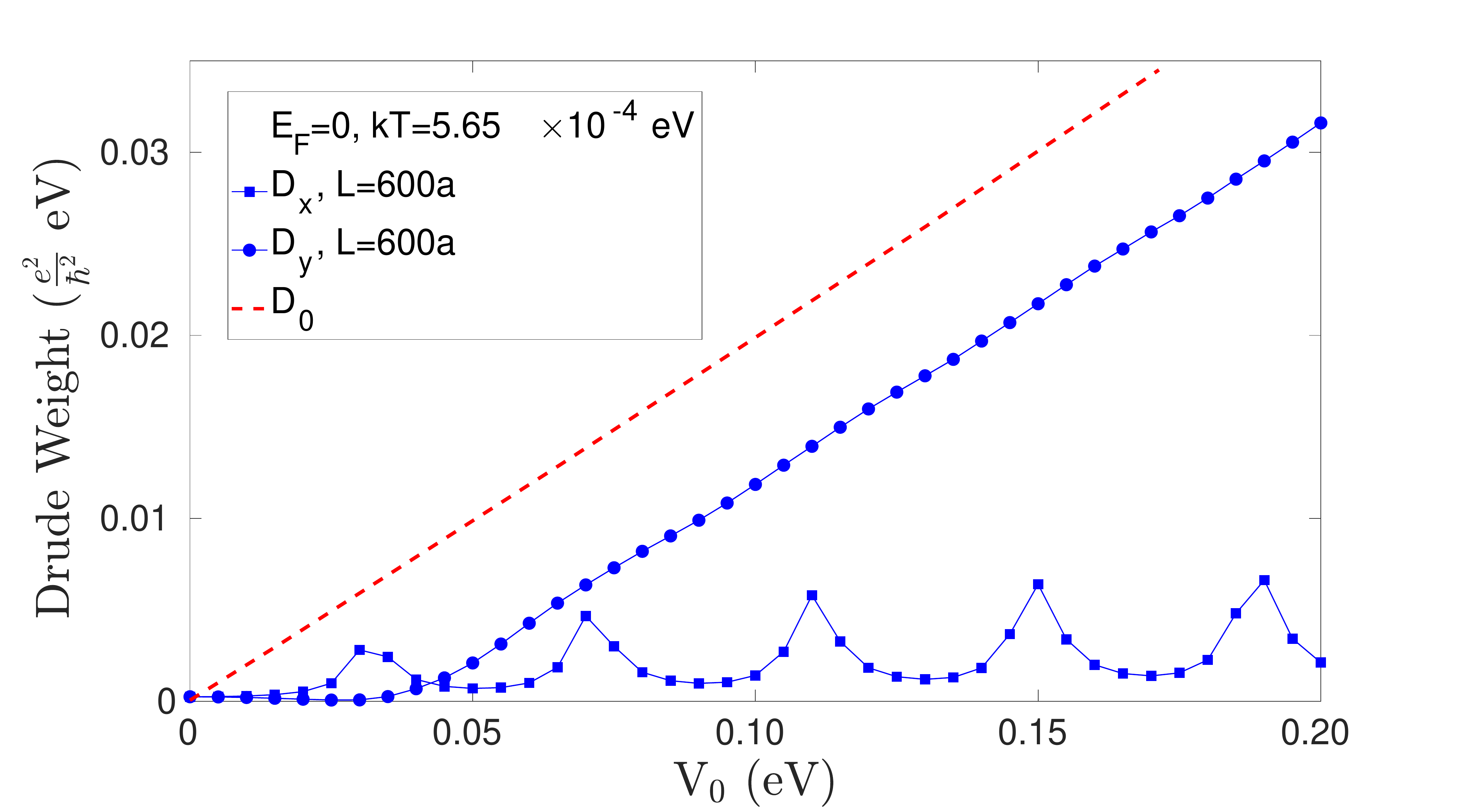}
\caption{}
\label{FigDrudekT}
\end{subfigure}
\begin{subfigure}{.45\textwidth}
\includegraphics[width=8cm]{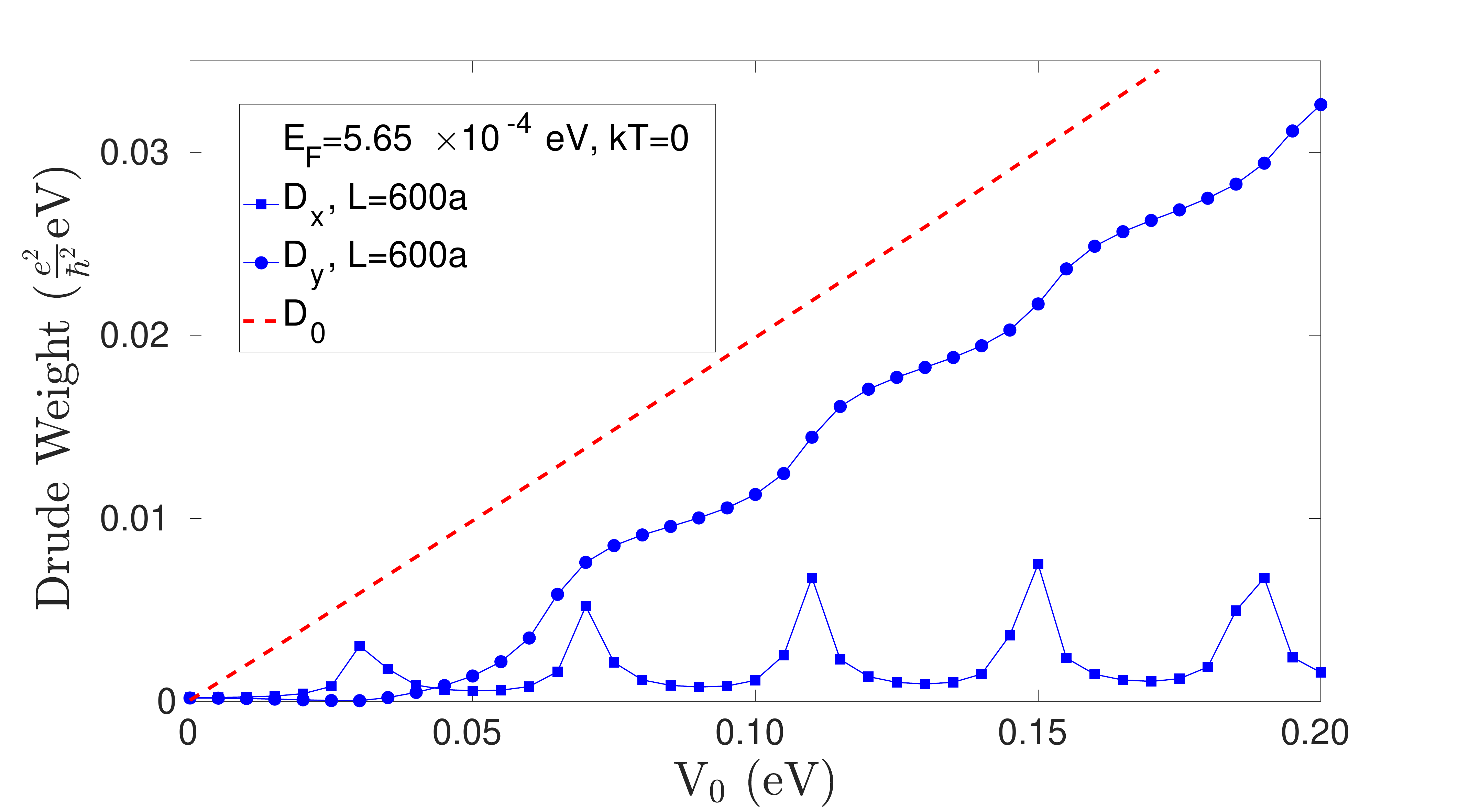}
\caption{}  
\label{FigDrudemu}
\end{subfigure}
\caption{Drude weight calculated from the band structure (Eqs. \ref{Dband} and \ref{T}) as function of the amplitude of the superlattice potential, $V_0$. (a): zero Fermi energy, $E_F=0\,$, and finite temperature, $kT=5.65 \times 10^{-4}\,$eV. (b): zero temperature, $kT=0\,$, and finite Fermi level, $E_F=5.65 \times 10^{-4}\,$eV. 
$\hat x$ is the direction of the SL axis and $\hat y$ is the perpendicular direction.  The period of the  superlattice is $L$=600$a$. The Drude weight corresponding to ${\overline E}_F=\frac{2}{\pi}V_0$ of a homogeneous system, i.e., $D_0$=$\frac {e^2}{\hbar^2} \frac{{\overline E}_F}{\pi}$, is shown as dashed line. All Drude weights are in units of $\frac {e^2}{\hbar^2}$eV. }
\end{figure} 
%

The results of this calculation for the neutral system, $E_F=0$, but at a finite temperature, $kT=5.65 \times 10^{-4}\,$eV, of the order of the damping in the original KK calculation, are shown in Fig. \ref{FigDrudekT}. Comparing Fig. \ref{FigDrudekT} with Fig. \ref{FigureDrudeN}, one is easily convinced of the physical correctness of our reasoning. Notice that temperature is not the only way to introduce a finite scale in  Eq. \ref{Dband}. For instance, we could also take zero temperature, but now choose a finite Fermi level.  The results of Eq. \ref{Dband} at zero temperature but  $E_F=5.65\times 10^{-4}\,$eV are shown in Fig. \ref{FigDrudemu}. Therefore, the physical effects of adding a finite energy scale to the neutral system, either in temperature or Fermi level, remain strikingly similar and, furthermore, show the same behavior as the corresponding KK calculation, where the role of the finite energy scale is taken by the finite damping.  

\section{Local approximations} 
Local approximations to the full linear response are widely used in the literature. Here, we summarise two approximations: the quantum mechanical, but local response (Q-local) and the semiclassical Thomas-Fermi approach.
\subsection{Q-local conductivity}
The optical conductivity tensor $ {\underline{\sigma} }({\bf r},{\bf r}')$ encodes the linear response of the electric current density to the electric field ${\bf E} ({\bf r})$ via 
\begin{equation}
{\bf J}({\bf r} )= \int d{\bf r}' \underline {\sigma} ({\bf r},{\bf r}'){\bf E} ({\bf r}')\;.
\label{opt-c}
\end{equation}
Te simplify notation, we have neglected the frequency dependence. In reciprocal space, the previous relation reads
\begin{equation}
{\bf J} ({\bf q})=\frac 1 {S} \sum _{{\bf q}'} \underline  \sigma ({\bf q}, {\bf q}') {\bf E}({\bf q}') \, \, ,
\end{equation}
with $S$ being the sample dimension and the Fourier transform defined as
\begin{equation}
{\underline \sigma}({\bf q},{\bf q}')=\int d{\bf r} \,  d{\bf r}' e^{i {\bf q}{\bf r}} e^{-i{\bf q}' \, {\bf r}'} {\underline \sigma}({\bf r},{\bf r}')\;.
\end{equation}

The local approximation assumes that the electrical current at point ${\bf r}$ only depends on the value of the electric fields at points
${\bf r}'$=${\bf r} +{\bf R}$ near ${\bf r}$, i.e., for values of ${\bf R}$ near zero.  Then, making an expansion for ${\bf R} \rightarrow$0 which is valid assuming that ${\underline \sigma}$ is strongly peaked around ${\bf R}=0$, we get
\begin{eqnarray}
{\underline \sigma}({\bf q},{\bf q}')&= & \! \! \int d{\bf R} d{\bf r} e^{i ({\bf q}-{\bf q}'){\bf r}} e^{-i{\bf q}' \, {\bf R}} {\underline \sigma}({\bf r},{\bf r}+{\bf R}) \nonumber \\
&=&\! \! \int  d{\bf r} e^{i ({\bf q}-{\bf q}'){\bf r}}   \! \! \int   d{\bf R}  (1-i{\bf q}' \, {\bf R}+ ...) {\underline \sigma}({\bf r},{\bf r}+{\bf R}) \nonumber \\
&\approx& \! \! \int  d{\bf r} e^{i ({\bf q}-{\bf q}'){\bf r}}   \! \! \int   d{\bf R} \, {\underline \sigma}({\bf r},{\bf r}+{\bf R}) \nonumber \\
&=&\! \!  \int  d{\bf r} e^{i ({\bf q}-{\bf q}'){\bf r}}  \! \!  \int   d{\bf R} \, {\underline \sigma}({\bf r},{\bf R}) \nonumber \\
&=& \! \! {\underline\sigma}({\bf q}-{\bf q}',0)
\end{eqnarray}
Restoring the notation of the main text, we thus have
\begin{equation}
\sigma^{\nu \nu} ({\bf q}+G, {\bf q}+ G';  \omega) \xrightarrow[\text{local}]{} \sigma^{\nu \nu} (G-G',0;  \omega)\equiv\sigma_L^{\nu \nu}(G-G';\omega)\, \, .
\label{local}
\end{equation}
\subsection{Semiclassical local approach}
The plasmon dispersion of a graphene sheet with a spatial modulation of the Drude weight has be obtained within a local response formalism \cite{Peres:2012aa,Slipchenko:2013aa,silveiro:2013aa,Beckerleg:2016aa,Huidobro:2016aa,Huidobro:2016ab,nikitin-book}. Let us here recall the basic steps to make the discussion self-contained. The relation of the local approach to microscopic calculations is also elucidated.
\subsubsection{Local Drude response}
We assume a metallic sheet with a spatially modulated Drude response $D(\bm r)$ so that the electron current obeys
\begin{equation}\label{drude}
\partial_t \bm j(\bm r,t) = D(\bm r) \bm E(\bm r,t)
,\end{equation}
where $\bm E = - \bm \nabla \phi$ is the electric field derived from the potential $ \phi $. The constitutive Eq. (\ref{drude}) encodes the material response. Notice that, even if $D$ were spatially constant, Eq. (\ref{drude}) still implies a local response approximation.

Assuming all time dependencies as $\exp(-i \omega t) $ and making use of the continuity equation, one obtains the following basic relation between charge and potential
\begin{equation}\label{basic}
\omega^2 \rho(\bm r) = -\bm \nabla \cdot D(\bm r) \bm \nabla \phi(\bm r)
.\end{equation}
We consider a periodic modulation
\begin{equation}\label{Dcrystal}
 D(\bm r) = \sum_{\bm G} d_{\bm G} \;\ee^{i \bm G \cdot \bm r}
,\end{equation}
where the sum runs over all reciprocal lattice vectors of the periodic structure. Writing the potential in  Bloch-like form 
\begin{equation}\label{phi}
 \phi (\bm r) = \ee^{i \bm q \cdot \bm r} \sum_{\bm G} a_{\bm G} \; \ee^{i \bm G \cdot \bm r}
,\end{equation}
and making use of the relation between Fourier components of charge and potential, $\rho_{\bm q} = 2 \epsilon_0 (\bm q \cdot \bm q)^{1/2} \phi_{\bm q} $, the following generalized eigenvalue equation is obtained
\begin{equation}\label{eigen1}
 2 \epsilon_0 \omega^2  |\bm q + \bm G|
\delta_{\bm G,\bm G'} \, a_{\bm G'} =\mathcal{M}_{\bm G,\bm G'}\, a_{\bm G'} 
,\end{equation}
where $|\bm q + \bm G| = [(\bm q + \bm G)\cdot(\bm q + \bm G)]^{1/2}$ and implied summation over repeated $ \bm G'$. The matrix $\mathcal{M} $ is given by 
\begin{equation}\label{M}
 \mathcal{M}_{\bm G,\bm G'} = (\bm q + \bm G)\cdot(\bm q + \bm G') d_{\bm G-\bm G'}
.\end{equation}
 Using the transformation 
$\tilde{a}_{\bm G} = \mathcal{S}_{\bm G,\bm G'} a_{\bm G'} $, where $\mathcal{S}_{\bm G,\bm G'}=  |\bm q + \bm G| ^{1/2} \; \delta_{\bm G,\bm G'}$,
Eq. (\ref{eigen1}) can be cast in the traditional eigenvalue form: 
\begin{equation}\label{eigen2}
 2 \epsilon_0 \omega^2 
\delta_{\bm G,\bm G'} \, \tilde{a}_{\bm G'} =\tilde{\mathcal{M}}_{\bm G,\bm G'}\, \tilde{a}_{\bm G'} 
,\end{equation}
with 
\begin{equation}\label{tildeM}
 \tilde{\mathcal{M}}_{\bm G,\bm G'} = |\bm q + \bm G|^{-1/2} \mathcal{M}_{\bm G,\bm G'} |\bm q + \bm G'|^{-1/2}
.\end{equation}

For a one-dimensional modulation along the $x$-direction of period $a$ one has
\begin{equation}\label{ds}
D(x) = d_0 + \sum_{n \neq 0} d_n \ee^{i n G_0 x} 
,\end{equation}
with $G_0=\tfrac{2 \pi }{L} $ and $d_n = d^*_{-n} $. Using dimensionless variables $\tilde{q}_{x,y} = \tfrac{q_{x,y}}{Q_0} $, $\tilde{d}_n= \tfrac{d_n}{d_0} $ and $\tilde{\omega} = \tfrac{\omega}{\omega_0} $, with  $\omega_0^2 = \tfrac{d_0 \, G_0}{2 \epsilon_0} $, the eigenvalue equation reads  
\begin{equation}\label{eigen3}
\tilde{\omega}^2 \;  \delta_{n,n'}\tilde{a}_{n'} =\tilde{\tilde{\mathcal{M}}}_{n,n'} \, \tilde{a}_{n'}  
,\end{equation}
with
\begin{equation}\label{2tildeM}
\tilde{\tilde{\mathcal{M}}}_{n,n'} = \frac{(n+\tilde{q}_x)(n'+\tilde{q}_x) + \tilde{q}_y^2}
{[(n+\tilde{q}_x)^2+\tilde{q}_y^2]^{1/4} \; [(n'+\tilde{q}_x)^2+\tilde{q}_y^2]^{1/4}} \; \tilde{d}_{n-n'}
.\end{equation}

The metal properties are contained in the coefficients $ \tilde{d}_{n} $. Notice that, if one assumes  just one  off-diagonal term treated as a perturbation
 $(\tilde{d}_{n=\pm 1} <<1)$, the plasmon frequencies for $q_{x,y}=0 $ do not change to lowest order. In this limit, such plasmon frequencies are those  of a  homogeneous metal with the averaged Drude weight $(d_0)$ for the wave-vector $ n G_0$,  with $n=0,\pm 1, \pm 2 ... $, though of course, they necessarily appear as {\it zone-folded} at the center of the Brillouin zone.

For graphene, assuming the modulation parameterized by $k_F(x) $, the (local) Fermi wavevector,
\begin{equation}\label{k_F}
k_F(x) = k_0 + \sum_{n \neq 0} k_n \ee^{i n G_0 x} 
,\end{equation}
the coefficients $ \tilde{d}_{n}  $ can be obtained under the assumption that the proportionality between Drude and Fermi wavevector also holds locally, again a kind of second instance of local approximation, 
\begin{equation}\label{Dkf}
D(x) \propto  |k_F(x)| 
.\end{equation}
Notice that the Drude weight of homogeneous graphene does not depend on the sign of carriers. Therefore,  a spatial modulation in $k_F(x)$ of period $a$ in {\it undoped} graphene will imply a modulation in $D(x)$ of period $a/2$.

 \subsubsection{Relation to Microscopic Calculation.}
The basic ingredient of a calculation \`a la RPA is the charge response  of the non-interacting {\it microscopic} Hamiltonian to a potential {\it (bubble)}. Writing the potential as 
\begin{equation}\label{cac1}
\phi(\bm r,t) = \phi_{\bm G'}(\bm q,\omega) \ee^{i (\bm q + \bm G') \cdot \bm r}\ee^{-i \omega t}
,\end{equation}
and the charge response as
\begin{equation}\label{cac2}
\rho(\bm r,t) = \sum_{\bm G} \rho_{\bm G}(\bm q,\omega) \ee^{i (\bm q + \bm G) \cdot \bm r}\ee^{-i \omega t}
,\end{equation}
then
\begin{equation}\label{cac3}
\rho_{\bm G}(\bm q,\omega) = \chi^o_{\bm G,\bm G'} (\bm q,\omega) \; \phi_{\bm G'}(\bm q,\omega)
,\end{equation}
where $ \chi^o_{\bm G,\bm G'}$ corresponds to the usual non-interacting bubble calculation. The charge response is related to  the conductivity used in the numerical calculations by $ \chi^o_{\bm G,\bm G'} (\bm q,\omega) = \tfrac{1}{i \omega} (q_{\alpha} +  G_{\alpha}) \sigma^{\alpha \beta}_{\bm G,\bm G'} (\bm q,\omega) (q_{\beta} + G'_{\beta})$.

 As the tenet of the RPA approximation is that  $ \chi^o$ is also the response of the real system (with Coulomb interaction) to the total potential (external plus Coulomb induced), then it can be obtained directly from Eq. (\ref{eigen1}) and Eq. (\ref{M}) as
\begin{equation}\label{chi}
\chi^o_{\bm G, \bm G'}(\bm q,\omega) = \frac{(\bm q + \bm G)\cdot(\bm q + \bm G')}{\omega^2} \, d_{\bm G-\bm G'}
.\end{equation}
In order to understand the meaning of Eq. (\ref{chi}), let us consider the case of a metal without modulation, where only $d_0 $ survives. One then has that Eq. (\ref{chi}) is equivalent to the charge-conservation requirement between charge and current responses,
\begin{equation}\label{req}
\omega^2 \chi^o = q^2 \chi_{jj}
,\end{equation}
provided we identify $d_0$ with the current correlation.  This is precisely the meaning of the Drude weight, of course, 
 understood as valid in the present context within the {\it metallic} limit:
\begin{equation}\label{limit}
d_0 = \lim_{\omega \to 0} \lim_{\q \to 0} \chi_{jj}(q,\omega)
.\end{equation}

Eq. (\ref{chi}) also allows us to understand the limits of the present local approach as compared with the  full microscopic calculation. For instance, if one wished to extract the $d_n$ from Eq. (\ref{chi}), they would end up being complex functions of $(\bm q,\omega) $, whereas in the local approach they are just (real, for x-even modulation) constants. Furthermore, $d_n$ in the local approach is the same number for all processes where the momentum changes by $n G_0 $ whereas in the full calculation, $ \chi^o_{\bm G, \bm G'}$ need not depend only on  the difference $\bm G- \bm G'$.

On the other hand, the local approach seems to easily capture features such as the role of the average Drude weight and the periodicity halving for neutral graphene.

\subsubsection{Semiclassical Thomas-Fermi conductivity}
The semiclassical Thomas-Fermi approach assumes that the system is slowly varying on the scale of the inverse Fermi wavenumber $k_F=\frac{E_F}{\hbar v_F}$. The system can then be described by a locally well-defined Fermi energy. Furthermore, at low energies and long distances, graphene can be considered homogeneous and isotropic. Its  electronic properties  are then well described by the Dirac equation and its response can be calculated analytically. In the local approximation, ${\bf q}$=0, the optical conductivity of graphene  depends only on frequency  and on  the Fermi energy with respect to the Dirac point and has the form\cite{Falkovsky07,Stauber08}  
\begin{equation}
\label{SemiclassicalLocal}
\sigma  (\omega, E_F)\! \! =  \! \! \sigma _0 \Theta(\hbar \omega-2E_F)+ i  \frac {\sigma_0}{\pi} \left ( \frac {4 E_F}{\hbar \omega}  -  \log \left | \frac {\hbar \omega -2 E_F}{\hbar \omega +2 E_F }\right | \right )\;.
\end{equation} 
where $\sigma _0$=$\frac 1 4 \frac {e^2} {\hbar}$.
 
The optical conductivity at each point in space is now determined by the Fermi energy at that point,  which in turn is obtained from the local charge density $n=\frac{k_F^2}{\pi}$ using the Thomas-Fermi approximation
\begin{equation}
\sigma(r)=\sigma(\omega,E_F[n(r)])\;.
\end{equation} 

\section{Period Halving in Modulated Neutral Graphene}
Here, we show that period halving is an exact symmetry for the charge response of neutral graphene for a large class of potential modulations. The  general electron-hole symmetry of the  response for any modulation is also  presented. 
\subsection{Period halving} 
Period halving is explained quite naturally for modulated neutral graphene in the local approach. This strongly suggests that it is an exact result. Here we show that this is the case for a general potential,
\begin{equation}\label{V0}
V(x) =  \sum_{n \, \text{odd}} v_n \; \ee^{i n G_0 x}\;, 
\end{equation}
with $G_0=\tfrac{2 \pi }{L} $, where  one has
\begin{equation}\label{V}
V(x + L/2) =  -V(x)\;.
\end{equation}
Writing the Hamiltonian  as 
\begin{equation}\label{H}
H = H_0 + V(x)\;,
\end{equation}
where $H_0$ corresponds to unperturbed graphene, and making use of  the unitary transformation
\begin{equation}\label{U}
 U = \sigma_z \exp(-\frac{i}{\hbar} \frac{L}{2} P_x)\;,
\end{equation}
where $ \sigma_z $ is the pseudospin Pauli matrix and $P_x$ is the momentum operator,  $H$ transforms as,
\begin{equation}\label{Ht}
  U H U^{\dagger} = -H\;.
\end{equation}
The transformation $U$ implements a half-period translation plus B-sublattice negation.  The eigenstates of the Hamiltonian, $H \,|i> = E_i \,|i>$, remain eigenstates under the transformation, $H \,|\tilde{i}> = E_{\tilde{i}} \,|\tilde{i}>$, with  $ |\tilde{i}> = U \, |i>$ and $E_{\tilde{i}} = -E_i $. Writing the charge response in the transformed eigenbasis,
\begin{equation}\label{Kubo}
\chi^o_{n G_0, n' G_0} (\bm q,\omega) = \sum_{i,j} \frac{< \tilde{i}| \rho_{\bm q + n G_0 \hat{\bm x}} | \tilde{j} >< \tilde{j}| \rho^{\dagger}_{\bm q + n' G_0 \hat{\bm x}} |\tilde{i}>}
{\omega+\ii \eta+(E_{\tilde{j}} - E_{\tilde{i}})} 
(n_F(E_{\tilde{i}}) - n_F(E_{\tilde{j}}) ) 
,\end{equation}
and making use of the following transformation property for the Fourier components of the charge density,
\begin{equation}\label{Fourier}
 U \, \rho_{\bm q + n G_0 \hat{\bm x}} \, U^{\dagger} = e^{-i q_x \tfrac{L}{2}} (-1)^n \rho_{\bm q + n G_0 \hat{\bm x}}
,\end{equation}
together with 
\begin{equation}\label{Fermirel}
 n_F(E_{\tilde{i}},\mu) - n_F(E_{\tilde{j}},\mu)  = n_F(E_j,-\mu) - n_F(E_i,-\mu)
,\end{equation}
where we have explicited the $\mu$ dependence of Fermi factors, then one easily arrives at the following symmetry of the charge-charge response, 
%
\begin{equation}\label{result}
\chi^o_{n G_0, n' G_0} (\bm q,\omega; V,\mu) =  (-1)^{n-n'}  \chi^o_{n G_0, n' G_0} (\bm q,\omega; V,-\mu)\;, 
\end{equation}
where time reversal symmetry has been also invoked. 

Notice that in Eq. (\ref{result}), one compares the response of the modulated graphene Hamiltonian, Eq. (\ref{H}), at opposite chemical potentials, $\pm \mu $. For neutral graphene one has $\mu=-\mu=0$ and, therefore, all odd terms vanish: 
\begin{equation}\label{odd}
\chi^o_{n G_0, n' G_0} = 0,  \qquad {\text for} \quad n-n'= {\rm odd}\;.
\end{equation}
This is the statement of period halving at charge neutrality for modulations of complying with Eq. \ref{V}, valid at any temperature. 
 
\subsection{General electron-hole response symmetry}

Using the above procedure but now for the transformation $U=\sigma_z$, one can show that 
\begin{equation}\label{general}
\chi^o_{n G_0, n' G_0} (\bm q,\omega; V,\mu) =    \chi^o_{n G_0, n' G_0} (\bm q,\omega; -V,-\mu)
,\end{equation}
valid for {\it any}  $V(x)$, irrespective of whether or not  Eq. (\ref{V}) is satisfied. Eq. (\ref{general}) represents the  general electron-hole symmetry of the graphene response for any modulation.

\section{Second sub-band plasmons at the centre of the Brillouin zone}

\begin{figure}[htbp]
\includegraphics[width=10cm]{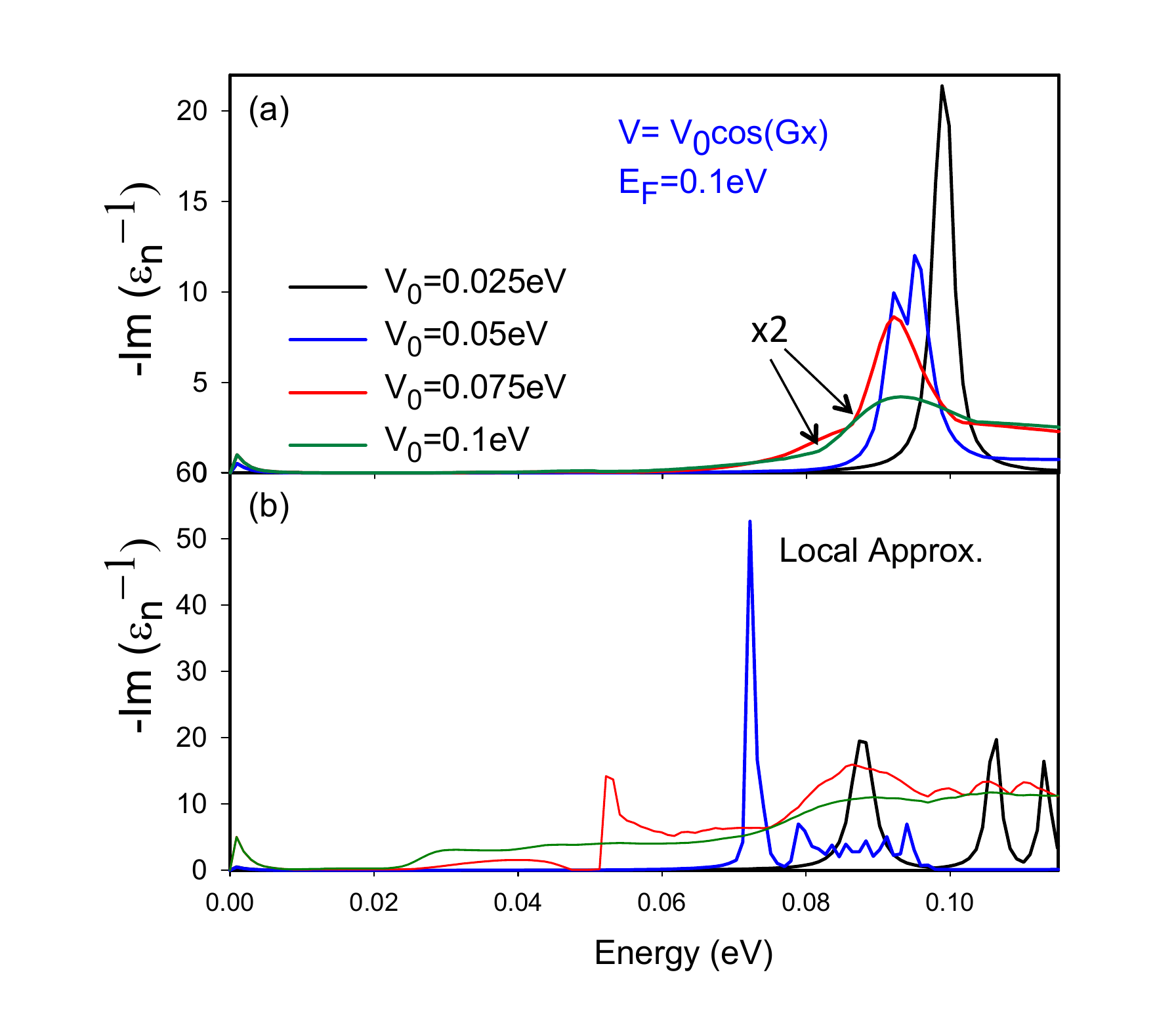}
\caption{Energy loss function at ${\bf q}$=0, for a SL with periodicity  $L$=600$a$, different values of  $V_0$ and  $E_F$=0.1eV. (b) The same but neglecting non-local effects in the
calculation (Q-local).}
\label{Figure2}
\end{figure} 

\begin{figure}[htbp]
\includegraphics[width=10cm]{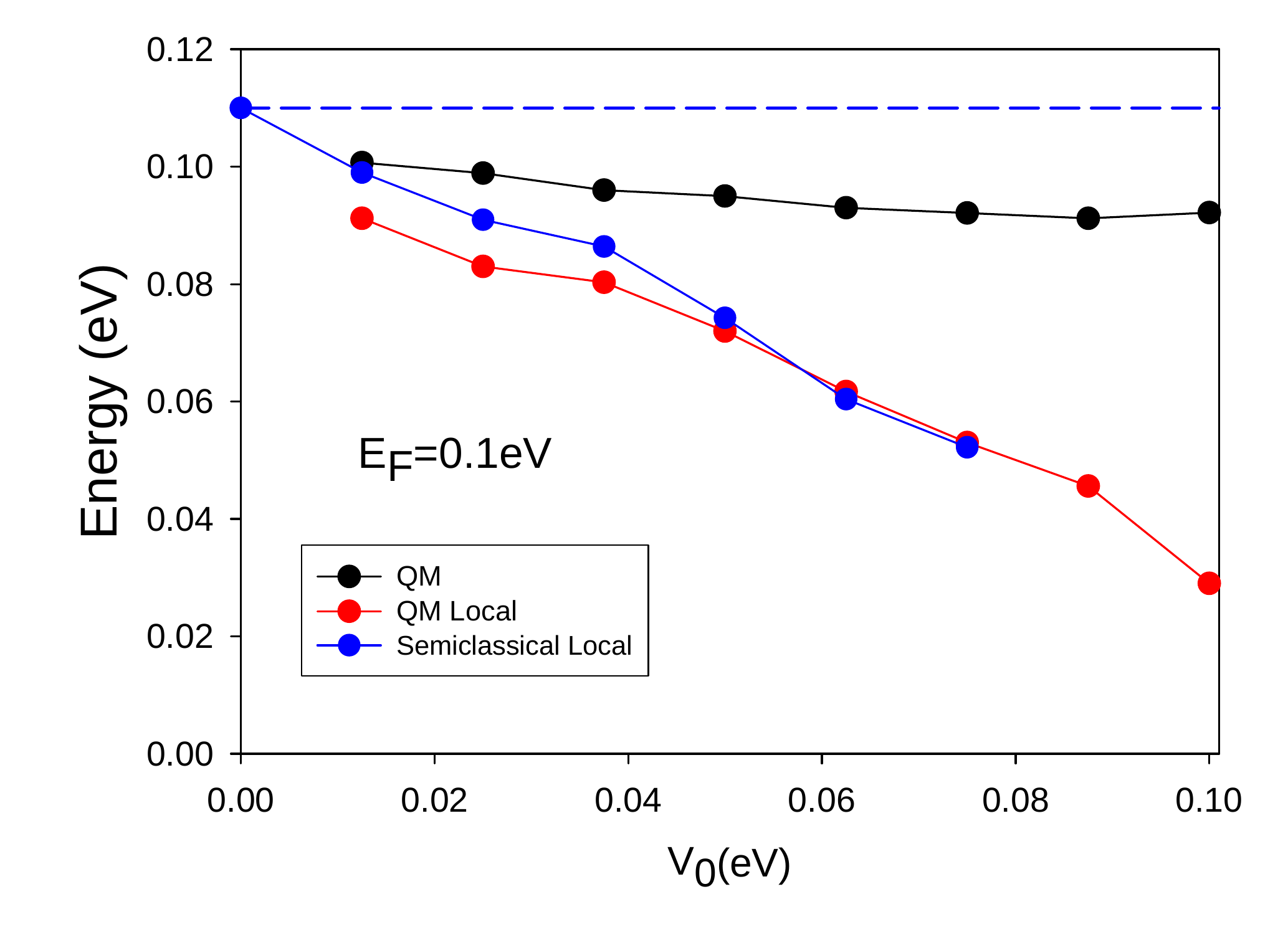}
\caption{
Energy of the main finite energy peak of the ${\bf q}$=0 energy loss function for different values of $V_0$ in the case of a SL potential with period $L$=600$a$ and doping level of $E_F$=0.1eV. The results correspond to three approximations, quantum mechanical non-local, quantum mechanical local and semiclassical local. The horizontal blue dashed line indicates the energy of a plasmon with momentum $G_0$ in pristine graphene in the Drude approximation and neglecting non-local effects.}
\label{Figure3}
\end{figure} 

At the center of the Brillouin zone, ${\bf q}$=0, the energy loss function shows strong peaks at finite energy. These modes correspond to electronic interband transitions that appear when the SL potential mixes electronic eigenstates states of the unperturbed graphene sheet with momentum $nG_0$.  For moderate values of $V_0$, the perturbation is rather weak and the $\q=0$ second sub-band plasmon should have an energy close to $\hbar \omega _p (G_0)$, with $\omega_p$ given in Eq. (1) of the main text. 

In Fig. \ref{Figure2}(a), we plot the energy loss function at the center of the Brillouin zone for a SL with period $L$=600$a$, different values of  $V_0$ and doping level of $E_F$=0.1eV.  For all values of $V_0$, there is a double peak structure that becomes wider and less intense for increasing $V_0$. Interestingly, the plasmonic resonance is even noticeable for $V_0$=$E_F$, for which there are points in space where the charge density becomes zero. At these so-called singular points, the semiclassical approximation breaks down.  

In Fig. \ref{Figure2}(b), we plot the energy loss function as obtained using the quantum local approximation for the optical conductivity of Eq. (\ref{local}). In this approach and for weak values of $V_0$, there are several finite energy peaks corresponding to higher harmonics of $G_0$. For moderate values of the SL amplitude, these peaks become very broad. Well defined collective excitations can thus not be identified, anymore.  

We summarise our results in Fig. \ref{Figure3} where we plot the energy of the strongest peak of the ${\bf q}$=0 energy loss function as function of $V_0$ for the quantum non-local, quantum local, and the semiclassical local approximations. In both local approximations, the presence of a SL modulation decreases the energy  of the plasmonic excitations and the corresponding electric fields become localised in low density/conductivity regions.

On the other hand, the energy of the plasmonic resonance in the non-local approximation is weakly dependent on $V_0$ and its energy is only slightly lower than the energy of the folded plasmons $\omega _p(G_0)$, Eq. (\ref{Plasmon}), which can be attributed to non-local effects. The weak dependence of the  plasmon peak on $V_0$ indicates that the electric field of the collective excitation is not confined to spatial regions with low density and the modes, i.e., the sine and cosine folding solutions are hardly modified by the SL perturbation. 

\section{Neutral second sub-band plasmons at the centre of the Brillouin zone}
Within our formalism, we can also treat the most extreme superlattice, an alternating $p$- and $n$-region. In Fig. \ref{Figure7}(a), we plot the energy regions of the modulated system as function of the position for which semi-classically plasmons may exits. In Fig. \ref{Figure7}(b), the energy loss function is plotted for the same SL which consists of two main modes indicated by the red lines. Interestingly, the main broad peak defining the plasmon excitation occurs in an energy window for which the semiclassical approximation predicts the plasmonic electric field should be expelled from large spatial regions. 

The nature of the collective interband excitations can be unravelled by analyzing the electric fields corresponding to the two main peaks in the full quantum non-local loss function, see Fig.\ref{Figure7}(c). The peaks have the form of a sine and cosine function with half the period compared to the period of the underlying SL. This indicates that plasmons depend only on the absolute values of the density of the carriers, independent of their hole or electron character.  The sine-like  and cosine-like form of the electric fields reflects that they originate from the folded plasmons of unperturbed graphene with momentum $\pm 2 G_0$.
\begin{figure}[htbp]
\includegraphics[width=10cm]{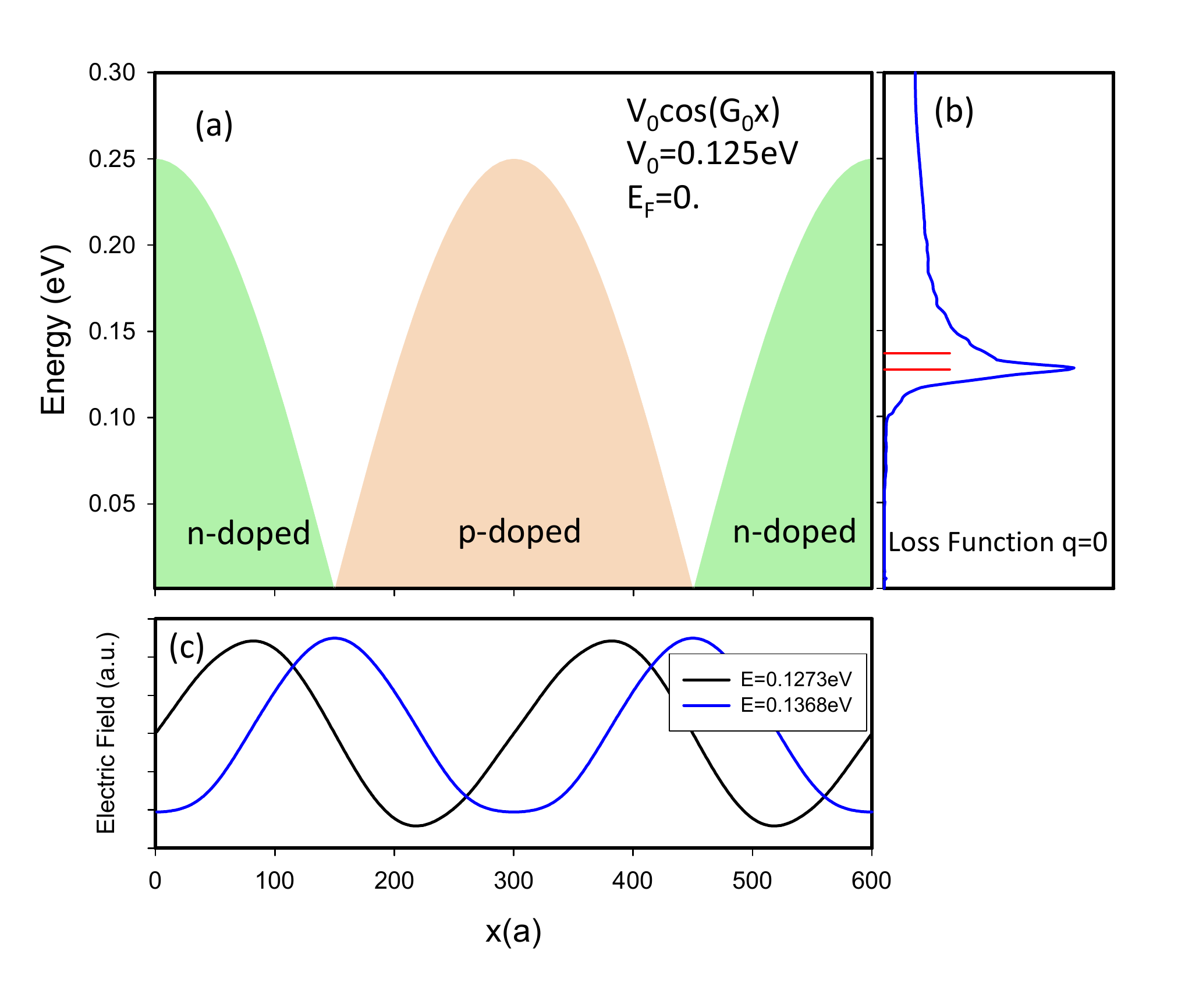}
\caption{(a) Schematic representation of a SL charge modulations of period $L$ and amplitude $V_0$=0.125eV, i.e., of the regions in the energy-position space where the imaginary part of $ \sigma _(|V_0 \cos G_0x)| $ is larger than zero and semiclassically, plasmons could be expected. These regions consists of $p$-doped and $n$-doped sectors.  In panel (b), we show the energy loss function for $q$=0 for the same SL parameters as in panel (a). A detailed analysis indicates that the large peak has really a double peak structure with energies E=0.1273eV and  E=0.1268eV.
Panel (c) shows the real  part of the electric fields corresponding to the two strongest folded interband dielectric eigenmodes with ${\bf q}$=0.} 
\label{Figure7}
\end{figure} 

In Fig. \ref{Figure6}, we plot the ${\bf q}$=0 energy loss  function for a superlattice of period $L=600a$ for different values of the SL amplitude $V_0$. For large values of the modulation $V_0$, there exits a peak in the loss function that indicates the existence of well defined plasmons. For small values of $V_0$, the peak becomes very broad and a continuous background at high energies appears. The energies corresponding to the peaks of Fig. \ref{Figure6} can well be fitted by Eq. (\ref{TwoDPlasmons}) after suitable rescaling which shall be discussed in the following. 

\begin{figure}[htbp]
\includegraphics[width=9cm]{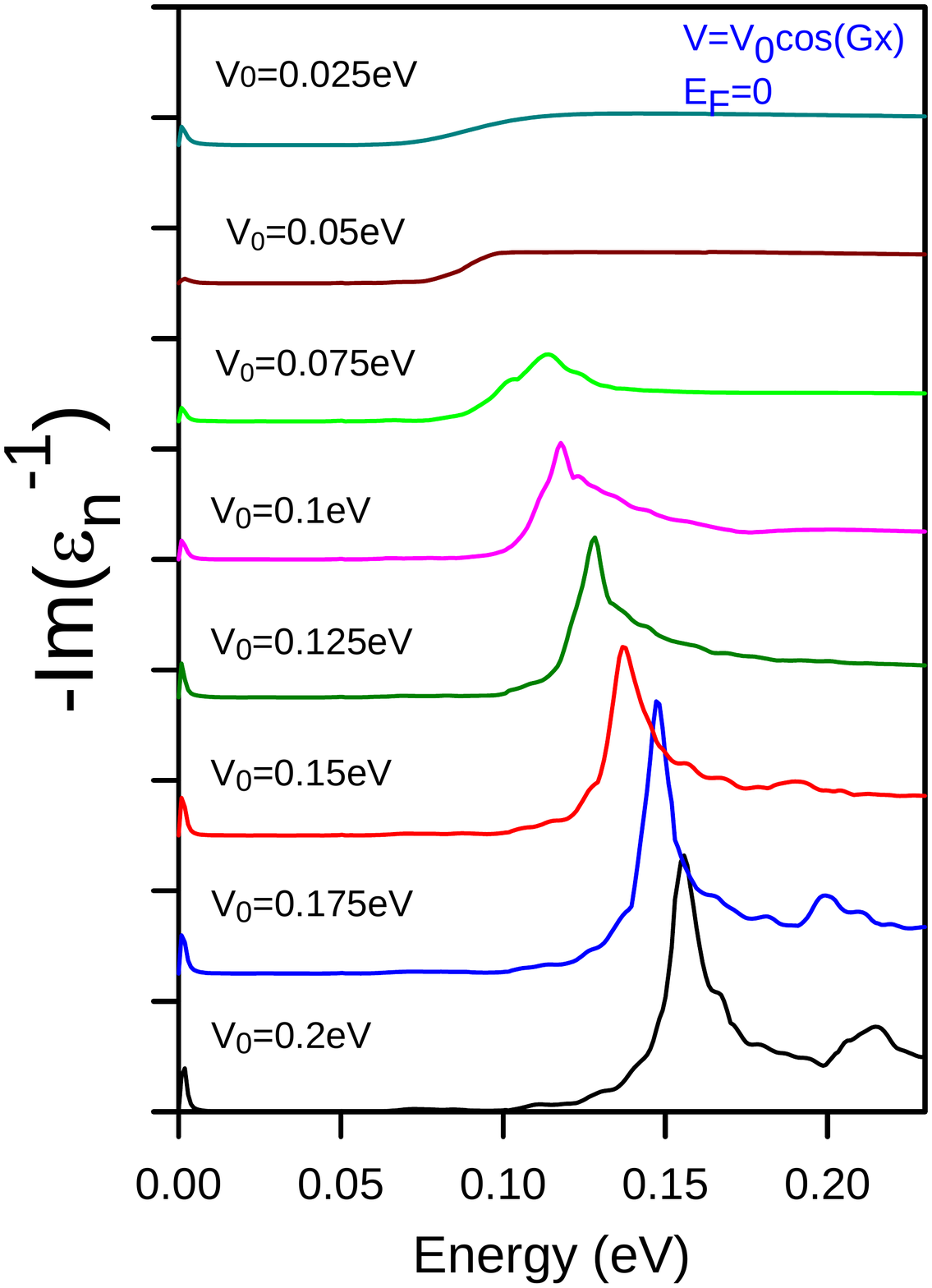}
\caption{Energy loss function evaluated at the center of the Brillouin zone for an undoped SL of period $L$=600$a$, for different values of $V_0\leq0.2$eV.} 
\label{Figure6}
\end{figure}

For neutral systems with modulation $V_0$, the effective Fermi energy is given by $\overline{E}_F=\frac{2}{\pi}V_0$. Plasmons at the zone-boundary and normalized by the effective Fermi energy thus correspond to 
\begin{equation}
\frac{\hbar\omega_p}{{\overline E}_F}=\sqrt{\frac{D\hbar}{2\varepsilon_0\epsilon v_F \overline{E}_F}\frac{2G_0}{\overline{k}_F}}\;.
\end{equation}
Using Eq. (\ref{localCurrentResponse}), the effective Drude weight is independent of $V_0$ and we have
\begin{equation}
\frac{D\hbar}{2\varepsilon_0\epsilon v_F \overline{E}_F}\to \frac{2\pi}{\epsilon}\alpha_g\;,
\end{equation}
where $\alpha_g\approx2.2$ denotes graphene's fine-structure.

Our numerical results yield two corrections. First, there is a non-local correction coming from the fact that the conductivity has a $\q$-dependence. Since the dispersion is obtained for fixed $G_0$, this correction is the same for all $V_0$ and thus simply leads to a renormalised Drude weight. Second, there is a red-shift originating from the level repulsion that occurs due to the superposition of the two plasmonic branches $\omega_p(\pm2G_0)$. This level repulsion is proportional to the SL potential $V_0$ and, therefore, also only leads to a renormalisation of the Drude weight. The plasmonic resonances can thus still be fitted by Eq. (\ref{TwoDPlasmons}). We interpret the decrease of Drude weight as an effective dipole-dipole interaction between the (semiclassically) separate conduction channels similar to the discussion in Ref. \cite{Strait13}.
\end{widetext}

\bibliography{mia}

\begin{thebibliography}{63}%
\makeatletter
\providecommand \@ifxundefined [1]{%
 \@ifx{#1\undefined}
}%
\providecommand \@ifnum [1]{%
 \ifnum #1\expandafter \@firstoftwo
 \else \expandafter \@secondoftwo
 \fi
}%
\providecommand \@ifx [1]{%
 \ifx #1\expandafter \@firstoftwo
 \else \expandafter \@secondoftwo
 \fi
}%
\providecommand \natexlab [1]{#1}%
\providecommand \enquote  [1]{``#1''}%
\providecommand \bibnamefont  [1]{#1}%
\providecommand \bibfnamefont [1]{#1}%
\providecommand \citenamefont [1]{#1}%
\providecommand \href@noop [0]{\@secondoftwo}%
\providecommand \href [0]{\begingroup \@sanitize@url \@href}%
\providecommand \@href[1]{\@@startlink{#1}\@@href}%
\providecommand \@@href[1]{\endgroup#1\@@endlink}%
\providecommand \@sanitize@url [0]{\catcode `\\12\catcode `\$12\catcode
  `\&12\catcode `\#12\catcode `\^12\catcode `\_12\catcode `\%12\relax}%
\providecommand \@@startlink[1]{}%
\providecommand \@@endlink[0]{}%
\providecommand \url  [0]{\begingroup\@sanitize@url \@url }%
\providecommand \@url [1]{\endgroup\@href {#1}{\urlprefix }}%
\providecommand \urlprefix  [0]{URL }%
\providecommand \Eprint [0]{\href }%
\providecommand \doibase [0]{http://dx.doi.org/}%
\providecommand \selectlanguage [0]{\@gobble}%
\providecommand \bibinfo  [0]{\@secondoftwo}%
\providecommand \bibfield  [0]{\@secondoftwo}%
\providecommand \translation [1]{[#1]}%
\providecommand \BibitemOpen [0]{}%
\providecommand \bibitemStop [0]{}%
\providecommand \bibitemNoStop [0]{.\EOS\space}%
\providecommand \EOS [0]{\spacefactor3000\relax}%
\providecommand \BibitemShut  [1]{\csname bibitem#1\endcsname}%
\let\auto@bib@innerbib\@empty
\bibitem [{\citenamefont {Ju}\ \emph {et~al.}(2011{\natexlab{a}})\citenamefont
  {Ju}, \citenamefont {Geng}, \citenamefont {Horng}, \citenamefont {Girit},
  \citenamefont {Martin}, \citenamefont {Hao}, \citenamefont {Bechtel},
  \citenamefont {Liang}, \citenamefont {Zettl}, \citenamefont {Shen},\ and\
  \citenamefont {Wang}}]{Ju11}%
  \BibitemOpen
  \bibfield  {author} {\bibinfo {author} {\bibfnamefont {L.}~\bibnamefont
  {Ju}}, \bibinfo {author} {\bibfnamefont {B.}~\bibnamefont {Geng}}, \bibinfo
  {author} {\bibfnamefont {J.}~\bibnamefont {Horng}}, \bibinfo {author}
  {\bibfnamefont {C.}~\bibnamefont {Girit}}, \bibinfo {author} {\bibfnamefont
  {M.}~\bibnamefont {Martin}}, \bibinfo {author} {\bibfnamefont
  {Z.}~\bibnamefont {Hao}}, \bibinfo {author} {\bibfnamefont {H.~A.}\
  \bibnamefont {Bechtel}}, \bibinfo {author} {\bibfnamefont {X.}~\bibnamefont
  {Liang}}, \bibinfo {author} {\bibfnamefont {A.}~\bibnamefont {Zettl}},
  \bibinfo {author} {\bibfnamefont {Y.~R.}\ \bibnamefont {Shen}}, \ and\
  \bibinfo {author} {\bibfnamefont {F.}~\bibnamefont {Wang}},\ }\href {\doibase
  10.1038/nnano.2011.146} {\bibfield  {journal} {\bibinfo  {journal} {Nature
  Nanotechnology}\ }\textbf {\bibinfo {volume} {6}},\ \bibinfo {pages} {630}
  (\bibinfo {year} {2011}{\natexlab{a}})}\BibitemShut {NoStop}%
\bibitem [{\citenamefont {Chen}\ \emph {et~al.}(2012)\citenamefont {Chen},
  \citenamefont {Badioli}, \citenamefont {Alonso-Gonzalez}, \citenamefont
  {Thongrattanasiri}, \citenamefont {Huth}, \citenamefont {Osmond},
  \citenamefont {Spasenovic}, \citenamefont {Centeno}, \citenamefont
  {Pesquera}, \citenamefont {Godignon}, \citenamefont {Zurutuza~Elorza},
  \citenamefont {Camara}, \citenamefont {de~Abajo}, \citenamefont
  {Hillenbrand},\ and\ \citenamefont {Koppens}}]{Chen12}%
  \BibitemOpen
  \bibfield  {author} {\bibinfo {author} {\bibfnamefont {J.}~\bibnamefont
  {Chen}}, \bibinfo {author} {\bibfnamefont {M.}~\bibnamefont {Badioli}},
  \bibinfo {author} {\bibfnamefont {P.}~\bibnamefont {Alonso-Gonzalez}},
  \bibinfo {author} {\bibfnamefont {S.}~\bibnamefont {Thongrattanasiri}},
  \bibinfo {author} {\bibfnamefont {F.}~\bibnamefont {Huth}}, \bibinfo {author}
  {\bibfnamefont {J.}~\bibnamefont {Osmond}}, \bibinfo {author} {\bibfnamefont
  {M.}~\bibnamefont {Spasenovic}}, \bibinfo {author} {\bibfnamefont
  {A.}~\bibnamefont {Centeno}}, \bibinfo {author} {\bibfnamefont
  {A.}~\bibnamefont {Pesquera}}, \bibinfo {author} {\bibfnamefont
  {P.}~\bibnamefont {Godignon}}, \bibinfo {author} {\bibfnamefont
  {A.}~\bibnamefont {Zurutuza~Elorza}}, \bibinfo {author} {\bibfnamefont
  {N.}~\bibnamefont {Camara}}, \bibinfo {author} {\bibfnamefont {F.~J.~G.}\
  \bibnamefont {de~Abajo}}, \bibinfo {author} {\bibfnamefont {R.}~\bibnamefont
  {Hillenbrand}}, \ and\ \bibinfo {author} {\bibfnamefont {F.~H.~L.}\
  \bibnamefont {Koppens}},\ }\href {\doibase 10.1038/nature11254} {\bibfield
  {journal} {\bibinfo  {journal} {Nature}\ }\textbf {\bibinfo {volume} {487}},\
  \bibinfo {pages} {77} (\bibinfo {year} {2012})}\BibitemShut {NoStop}%
\bibitem [{\citenamefont {Fei}\ \emph {et~al.}(2012)\citenamefont {Fei},
  \citenamefont {Rodin}, \citenamefont {Andreev}, \citenamefont {Bao},
  \citenamefont {McLeod}, \citenamefont {Wagner}, \citenamefont {Zhang},
  \citenamefont {Zhao}, \citenamefont {Thiemens}, \citenamefont {Dominguez},
  \citenamefont {Fogler}, \citenamefont {Neto}, \citenamefont {Lau},
  \citenamefont {Keilmann},\ and\ \citenamefont {Basov}}]{Fei12}%
  \BibitemOpen
  \bibfield  {author} {\bibinfo {author} {\bibfnamefont {Z.}~\bibnamefont
  {Fei}}, \bibinfo {author} {\bibfnamefont {A.~S.}\ \bibnamefont {Rodin}},
  \bibinfo {author} {\bibfnamefont {G.~O.}\ \bibnamefont {Andreev}}, \bibinfo
  {author} {\bibfnamefont {W.}~\bibnamefont {Bao}}, \bibinfo {author}
  {\bibfnamefont {A.~S.}\ \bibnamefont {McLeod}}, \bibinfo {author}
  {\bibfnamefont {M.}~\bibnamefont {Wagner}}, \bibinfo {author} {\bibfnamefont
  {L.~M.}\ \bibnamefont {Zhang}}, \bibinfo {author} {\bibfnamefont
  {Z.}~\bibnamefont {Zhao}}, \bibinfo {author} {\bibfnamefont {M.}~\bibnamefont
  {Thiemens}}, \bibinfo {author} {\bibfnamefont {G.}~\bibnamefont {Dominguez}},
  \bibinfo {author} {\bibfnamefont {M.~M.}\ \bibnamefont {Fogler}}, \bibinfo
  {author} {\bibfnamefont {A.~H.~C.}\ \bibnamefont {Neto}}, \bibinfo {author}
  {\bibfnamefont {C.~N.}\ \bibnamefont {Lau}}, \bibinfo {author} {\bibfnamefont
  {F.}~\bibnamefont {Keilmann}}, \ and\ \bibinfo {author} {\bibfnamefont
  {D.~N.}\ \bibnamefont {Basov}},\ }\href {\doibase 10.1038/nature11253}
  {\bibfield  {journal} {\bibinfo  {journal} {Nature}\ }\textbf {\bibinfo
  {volume} {487}},\ \bibinfo {pages} {82} (\bibinfo {year} {2012})}\BibitemShut
  {NoStop}%
\bibitem [{\citenamefont {Ni}\ \emph {et~al.}(2018)\citenamefont {Ni},
  \citenamefont {McLeod}, \citenamefont {Sun}, \citenamefont {Wang},
  \citenamefont {Xiong}, \citenamefont {Post}, \citenamefont {Sunku},
  \citenamefont {Jiang}, \citenamefont {Hone}, \citenamefont {Dean},
  \citenamefont {Fogler},\ and\ \citenamefont {Basov}}]{Ni18}%
  \BibitemOpen
  \bibfield  {author} {\bibinfo {author} {\bibfnamefont {G.~X.}\ \bibnamefont
  {Ni}}, \bibinfo {author} {\bibfnamefont {A.~S.}\ \bibnamefont {McLeod}},
  \bibinfo {author} {\bibfnamefont {Z.}~\bibnamefont {Sun}}, \bibinfo {author}
  {\bibfnamefont {L.}~\bibnamefont {Wang}}, \bibinfo {author} {\bibfnamefont
  {L.}~\bibnamefont {Xiong}}, \bibinfo {author} {\bibfnamefont {K.~W.}\
  \bibnamefont {Post}}, \bibinfo {author} {\bibfnamefont {S.~S.}\ \bibnamefont
  {Sunku}}, \bibinfo {author} {\bibfnamefont {B.~Y.}\ \bibnamefont {Jiang}},
  \bibinfo {author} {\bibfnamefont {J.}~\bibnamefont {Hone}}, \bibinfo {author}
  {\bibfnamefont {C.~R.}\ \bibnamefont {Dean}}, \bibinfo {author}
  {\bibfnamefont {M.~M.}\ \bibnamefont {Fogler}}, \ and\ \bibinfo {author}
  {\bibfnamefont {D.~N.}\ \bibnamefont {Basov}},\ }\href {\doibase
  10.1038/s41586-018-0136-9} {\bibfield  {journal} {\bibinfo  {journal}
  {Nature}\ }\textbf {\bibinfo {volume} {557}},\ \bibinfo {pages} {530}
  (\bibinfo {year} {2018})}\BibitemShut {NoStop}%
\bibitem [{\citenamefont {Alcaraz~Iranzo}\ \emph {et~al.}(2018)\citenamefont
  {Alcaraz~Iranzo}, \citenamefont {Nanot}, \citenamefont {Dias}, \citenamefont
  {Epstein}, \citenamefont {Peng}, \citenamefont {Efetov}, \citenamefont
  {Lundeberg}, \citenamefont {Parret}, \citenamefont {Osmond}, \citenamefont
  {Hong}, \citenamefont {Kong}, \citenamefont {Englund}, \citenamefont
  {Peres},\ and\ \citenamefont {Koppens}}]{Alcaraz18}%
  \BibitemOpen
  \bibfield  {author} {\bibinfo {author} {\bibfnamefont {D.}~\bibnamefont
  {Alcaraz~Iranzo}}, \bibinfo {author} {\bibfnamefont {S.}~\bibnamefont
  {Nanot}}, \bibinfo {author} {\bibfnamefont {E.~J.~C.}\ \bibnamefont {Dias}},
  \bibinfo {author} {\bibfnamefont {I.}~\bibnamefont {Epstein}}, \bibinfo
  {author} {\bibfnamefont {C.}~\bibnamefont {Peng}}, \bibinfo {author}
  {\bibfnamefont {D.~K.}\ \bibnamefont {Efetov}}, \bibinfo {author}
  {\bibfnamefont {M.~B.}\ \bibnamefont {Lundeberg}}, \bibinfo {author}
  {\bibfnamefont {R.}~\bibnamefont {Parret}}, \bibinfo {author} {\bibfnamefont
  {J.}~\bibnamefont {Osmond}}, \bibinfo {author} {\bibfnamefont {J.-Y.}\
  \bibnamefont {Hong}}, \bibinfo {author} {\bibfnamefont {J.}~\bibnamefont
  {Kong}}, \bibinfo {author} {\bibfnamefont {D.~R.}\ \bibnamefont {Englund}},
  \bibinfo {author} {\bibfnamefont {N.~M.~R.}\ \bibnamefont {Peres}}, \ and\
  \bibinfo {author} {\bibfnamefont {F.~H.~L.}\ \bibnamefont {Koppens}},\ }\href
  {\doibase 10.1126/science.aar8438} {\bibfield  {journal} {\bibinfo  {journal}
  {Science}\ }\textbf {\bibinfo {volume} {360}},\ \bibinfo {pages} {291}
  (\bibinfo {year} {2018})}\BibitemShut {NoStop}%
\bibitem [{\citenamefont {Nikitin}(2017)}]{nikitin-book}%
  \BibitemOpen
  \bibfield  {author} {\bibinfo {author} {\bibfnamefont {A.}~\bibnamefont
  {Nikitin}},\ }\href@noop {} {\emph {\bibinfo {title} {Graphene
  Plasmonics,}}},\ \bibinfo {series} {World Scientific Handbook of
  Metamateriasl and Plasmonics}, Vol.~\bibinfo {volume} {1}\ (\bibinfo
  {publisher} {Word Scientific Publishing},\ \bibinfo {year}
  {2017})\BibitemShut {NoStop}%
\bibitem [{\citenamefont {Koppens}\ \emph {et~al.}(2011)\citenamefont
  {Koppens}, \citenamefont {Chang},\ and\ \citenamefont {Garc{\'\i}a~de
  Abajo}}]{Koppens:2011aa}%
  \BibitemOpen
  \bibfield  {author} {\bibinfo {author} {\bibfnamefont {F.~H.~L.}\
  \bibnamefont {Koppens}}, \bibinfo {author} {\bibfnamefont {D.~E.}\
  \bibnamefont {Chang}}, \ and\ \bibinfo {author} {\bibfnamefont {F.~J.}\
  \bibnamefont {Garc{\'\i}a~de Abajo}},\ }\href {\doibase 10.1021/nl201771h}
  {\bibfield  {journal} {\bibinfo  {journal} {Nano Letters}\ }\textbf {\bibinfo
  {volume} {11}},\ \bibinfo {pages} {3370} (\bibinfo {year}
  {2011})}\BibitemShut {NoStop}%
\bibitem [{\citenamefont {Low}\ \emph {et~al.}(2016)\citenamefont {Low},
  \citenamefont {Chaves}, \citenamefont {Caldwell}, \citenamefont {Kumar},
  \citenamefont {Fang}, \citenamefont {Avouris}, \citenamefont {Heinz},
  \citenamefont {Guinea}, \citenamefont {Martin-Moreno},\ and\ \citenamefont
  {Koppens}}]{Low:2016aa}%
  \BibitemOpen
  \bibfield  {author} {\bibinfo {author} {\bibfnamefont {T.}~\bibnamefont
  {Low}}, \bibinfo {author} {\bibfnamefont {A.}~\bibnamefont {Chaves}},
  \bibinfo {author} {\bibfnamefont {J.~D.}\ \bibnamefont {Caldwell}}, \bibinfo
  {author} {\bibfnamefont {A.}~\bibnamefont {Kumar}}, \bibinfo {author}
  {\bibfnamefont {N.~X.}\ \bibnamefont {Fang}}, \bibinfo {author}
  {\bibfnamefont {P.}~\bibnamefont {Avouris}}, \bibinfo {author} {\bibfnamefont
  {T.~F.}\ \bibnamefont {Heinz}}, \bibinfo {author} {\bibfnamefont
  {F.}~\bibnamefont {Guinea}}, \bibinfo {author} {\bibfnamefont
  {L.}~\bibnamefont {Martin-Moreno}}, \ and\ \bibinfo {author} {\bibfnamefont
  {F.}~\bibnamefont {Koppens}},\ }\href {https://doi.org/10.1038/nmat4792}
  {\bibfield  {journal} {\bibinfo  {journal} {Nature Materials}\ }\textbf
  {\bibinfo {volume} {16}},\ \bibinfo {pages} {182 EP } (\bibinfo {year}
  {2016})}\BibitemShut {NoStop}%
\bibitem [{\citenamefont {Torbatian}\ and\ \citenamefont
  {Asgari}(2018)}]{Torbatian}%
  \BibitemOpen
  \bibfield  {author} {\bibinfo {author} {\bibfnamefont {Z.}~\bibnamefont
  {Torbatian}}\ and\ \bibinfo {author} {\bibfnamefont {R.}~\bibnamefont
  {Asgari}},\ }\href@noop {} {\bibfield  {journal} {\bibinfo  {journal}
  {Applied Science}\ }\textbf {\bibinfo {volume} {8}},\ \bibinfo {pages} {238}
  (\bibinfo {year} {2018})}\BibitemShut {NoStop}%
\bibitem [{\citenamefont {Garc{\'\i}a~de
  Abajo}(2014)}]{Garcia-de-Abajo:2014aa}%
  \BibitemOpen
  \bibfield  {author} {\bibinfo {author} {\bibfnamefont {F.~J.}\ \bibnamefont
  {Garc{\'\i}a~de Abajo}},\ }\href {\doibase 10.1021/ph400147y} {\bibfield
  {journal} {\bibinfo  {journal} {ACS Photonics}\ }\textbf {\bibinfo {volume}
  {1}},\ \bibinfo {pages} {135} (\bibinfo {year} {2014})}\BibitemShut {NoStop}%
\bibitem [{\citenamefont {Goncalves}\ and\ \citenamefont
  {Peres}(2016)}]{Peres-book}%
  \BibitemOpen
  \bibfield  {author} {\bibinfo {author} {\bibfnamefont {P.}~\bibnamefont
  {Goncalves}}\ and\ \bibinfo {author} {\bibfnamefont {N.}~\bibnamefont
  {Peres}},\ }\href@noop {} {\emph {\bibinfo {title} {An Introduction to
  Graphene Plasmonics}}}\ (\bibinfo  {publisher} {World Scientific, Singapur},\
  \bibinfo {year} {2016})\BibitemShut {NoStop}%
\bibitem [{\citenamefont {Bonaccorso}\ \emph {et~al.}(2010)\citenamefont
  {Bonaccorso}, \citenamefont {Sun}, \citenamefont {Hasan},\ and\ \citenamefont
  {Ferrari}}]{Bonaccorso:2010aa}%
  \BibitemOpen
  \bibfield  {author} {\bibinfo {author} {\bibfnamefont {F.}~\bibnamefont
  {Bonaccorso}}, \bibinfo {author} {\bibfnamefont {Z.}~\bibnamefont {Sun}},
  \bibinfo {author} {\bibfnamefont {T.}~\bibnamefont {Hasan}}, \ and\ \bibinfo
  {author} {\bibfnamefont {A.~C.}\ \bibnamefont {Ferrari}},\ }\href
  {https://doi.org/10.1038/nphoton.2010.186} {\bibfield  {journal} {\bibinfo
  {journal} {Nature Photonics}\ }\textbf {\bibinfo {volume} {4}},\ \bibinfo
  {pages} {611 EP } (\bibinfo {year} {2010})}\BibitemShut {NoStop}%
\bibitem [{\citenamefont {Farmer}\ \emph {et~al.}(2015)\citenamefont {Farmer},
  \citenamefont {Rodrigo}, \citenamefont {Low},\ and\ \citenamefont
  {Avouris}}]{Farmer:2015aa}%
  \BibitemOpen
  \bibfield  {author} {\bibinfo {author} {\bibfnamefont {D.~B.}\ \bibnamefont
  {Farmer}}, \bibinfo {author} {\bibfnamefont {D.}~\bibnamefont {Rodrigo}},
  \bibinfo {author} {\bibfnamefont {T.}~\bibnamefont {Low}}, \ and\ \bibinfo
  {author} {\bibfnamefont {P.}~\bibnamefont {Avouris}},\ }\href {\doibase
  10.1021/acs.nanolett.5b00148} {\bibfield  {journal} {\bibinfo  {journal}
  {Nano Letters}\ }\textbf {\bibinfo {volume} {15}},\ \bibinfo {pages} {2582}
  (\bibinfo {year} {2015})}\BibitemShut {NoStop}%
\bibitem [{\citenamefont {Fern{\'a}ndez-Dom{\'\i}nguez}\ \emph
  {et~al.}(2017)\citenamefont {Fern{\'a}ndez-Dom{\'\i}nguez}, \citenamefont
  {Garc{\'\i}a-Vidal},\ and\ \citenamefont
  {Mart{\'\i}n-Moreno}}]{Fernandez-Dominguez:2017aa}%
  \BibitemOpen
  \bibfield  {author} {\bibinfo {author} {\bibfnamefont {A.~I.}\ \bibnamefont
  {Fern{\'a}ndez-Dom{\'\i}nguez}}, \bibinfo {author} {\bibfnamefont {F.~J.}\
  \bibnamefont {Garc{\'\i}a-Vidal}}, \ and\ \bibinfo {author} {\bibfnamefont
  {L.}~\bibnamefont {Mart{\'\i}n-Moreno}},\ }\href
  {https://doi.org/10.1038/nphoton.2016.258} {\bibfield  {journal} {\bibinfo
  {journal} {Nature Photonics}\ }\textbf {\bibinfo {volume} {11}},\ \bibinfo
  {pages} {8 EP } (\bibinfo {year} {2017})}\BibitemShut {NoStop}%
\bibitem [{\citenamefont {Galiffi}\ \emph {et~al.}(2018)\citenamefont
  {Galiffi}, \citenamefont {Pendry},\ and\ \citenamefont
  {Huidobro}}]{Galiffi18}%
  \BibitemOpen
  \bibfield  {author} {\bibinfo {author} {\bibfnamefont {E.}~\bibnamefont
  {Galiffi}}, \bibinfo {author} {\bibfnamefont {J.~B.}\ \bibnamefont {Pendry}},
  \ and\ \bibinfo {author} {\bibfnamefont {P.~A.}\ \bibnamefont {Huidobro}},\
  }\href {\doibase 10.1021/acsnano.7b07951} {\bibfield  {journal} {\bibinfo
  {journal} {ACS Nano}\ }\textbf {\bibinfo {volume} {12}},\ \bibinfo {pages}
  {1006} (\bibinfo {year} {2018})}\BibitemShut {NoStop}%
\bibitem [{\citenamefont {Fan}\ \emph {et~al.}()\citenamefont {Fan},
  \citenamefont {Shen}, \citenamefont {Zhang}, \citenamefont {Zhao},
  \citenamefont {Wu}, \citenamefont {Fu}, \citenamefont {Wei}, \citenamefont
  {Li},\ and\ \citenamefont {Soukoulis}}]{Fan19}%
  \BibitemOpen
  \bibfield  {author} {\bibinfo {author} {\bibfnamefont {Y.}~\bibnamefont
  {Fan}}, \bibinfo {author} {\bibfnamefont {N.-H.}\ \bibnamefont {Shen}},
  \bibinfo {author} {\bibfnamefont {F.}~\bibnamefont {Zhang}}, \bibinfo
  {author} {\bibfnamefont {Q.}~\bibnamefont {Zhao}}, \bibinfo {author}
  {\bibfnamefont {H.}~\bibnamefont {Wu}}, \bibinfo {author} {\bibfnamefont
  {Q.}~\bibnamefont {Fu}}, \bibinfo {author} {\bibfnamefont {Z.}~\bibnamefont
  {Wei}}, \bibinfo {author} {\bibfnamefont {H.}~\bibnamefont {Li}}, \ and\
  \bibinfo {author} {\bibfnamefont {C.~M.}\ \bibnamefont {Soukoulis}},\
  }\href@noop {} {\bibfield  {journal} {\bibinfo  {journal} {Advanced Optical
  Materials}\ }\textbf {\bibinfo {volume} {7}},\ \bibinfo {pages}
  {1800537}}\BibitemShut {NoStop}%
\bibitem [{\citenamefont {Foerster}\ \emph {et~al.}(2019)\citenamefont
  {Foerster}, \citenamefont {Spata}, \citenamefont {Carter}, \citenamefont
  {S{\"o}nnichsen},\ and\ \citenamefont {Link}}]{Foerster19}%
  \BibitemOpen
  \bibfield  {author} {\bibinfo {author} {\bibfnamefont {B.}~\bibnamefont
  {Foerster}}, \bibinfo {author} {\bibfnamefont {V.~A.}\ \bibnamefont {Spata}},
  \bibinfo {author} {\bibfnamefont {E.~A.}\ \bibnamefont {Carter}}, \bibinfo
  {author} {\bibfnamefont {C.}~\bibnamefont {S{\"o}nnichsen}}, \ and\ \bibinfo
  {author} {\bibfnamefont {S.}~\bibnamefont {Link}},\ }\href {\doibase
  10.1126/sciadv.aav0704} {\bibfield  {journal} {\bibinfo  {journal} {Science
  advances}\ }\textbf {\bibinfo {volume} {5}},\ \bibinfo {pages} {eaav0704}
  (\bibinfo {year} {2019})}\BibitemShut {NoStop}%
\bibitem [{\citenamefont {Wunsch}\ \emph {et~al.}(2006)\citenamefont {Wunsch},
  \citenamefont {Stauber}, \citenamefont {Sols},\ and\ \citenamefont
  {Guinea}}]{Wunsch:2006aa}%
  \BibitemOpen
  \bibfield  {author} {\bibinfo {author} {\bibfnamefont {B.}~\bibnamefont
  {Wunsch}}, \bibinfo {author} {\bibfnamefont {T.}~\bibnamefont {Stauber}},
  \bibinfo {author} {\bibfnamefont {F.}~\bibnamefont {Sols}}, \ and\ \bibinfo
  {author} {\bibfnamefont {F.}~\bibnamefont {Guinea}},\ }\href {\doibase
  10.1088/1367-2630/8/12/318} {\bibfield  {journal} {\bibinfo  {journal} {New
  Journal of Physics}\ }\textbf {\bibinfo {volume} {8}},\ \bibinfo {pages}
  {318} (\bibinfo {year} {2006})}\BibitemShut {NoStop}%
\bibitem [{\citenamefont {Ando}(2006)}]{Ando:2006aa}%
  \BibitemOpen
  \bibfield  {author} {\bibinfo {author} {\bibfnamefont {T.}~\bibnamefont
  {Ando}},\ }\href {\doibase 10.1143/JPSJ.75.074716} {\bibfield  {journal}
  {\bibinfo  {journal} {Journal of the Physical Society of Japan}\ }\textbf
  {\bibinfo {volume} {75}},\ \bibinfo {pages} {074716} (\bibinfo {year}
  {2006})}\BibitemShut {NoStop}%
\bibitem [{\citenamefont {Hwang}\ and\ \citenamefont
  {Das~Sarma}(2007)}]{Hwang:2007aa}%
  \BibitemOpen
  \bibfield  {author} {\bibinfo {author} {\bibfnamefont {E.~H.}\ \bibnamefont
  {Hwang}}\ and\ \bibinfo {author} {\bibfnamefont {S.}~\bibnamefont
  {Das~Sarma}},\ }\href {\doibase 10.1103/PhysRevB.75.205418} {\bibfield
  {journal} {\bibinfo  {journal} {Physical Review B}\ }\textbf {\bibinfo
  {volume} {75}},\ \bibinfo {pages} {205418} (\bibinfo {year}
  {2007})}\BibitemShut {NoStop}%
\bibitem [{\citenamefont {Ando}\ \emph {et~al.}(1982)\citenamefont {Ando},
  \citenamefont {Fowler},\ and\ \citenamefont {Stern}}]{Ando:1982aa}%
  \BibitemOpen
  \bibfield  {author} {\bibinfo {author} {\bibfnamefont {T.}~\bibnamefont
  {Ando}}, \bibinfo {author} {\bibfnamefont {A.~B.}\ \bibnamefont {Fowler}}, \
  and\ \bibinfo {author} {\bibfnamefont {F.}~\bibnamefont {Stern}},\ }\href
  {\doibase 10.1103/RevModPhys.54.437} {\bibfield  {journal} {\bibinfo
  {journal} {Reviews of Modern Physics}\ }\textbf {\bibinfo {volume} {54}},\
  \bibinfo {pages} {437} (\bibinfo {year} {1982})}\BibitemShut {NoStop}%
\bibitem [{\citenamefont {Stauber}(2014)}]{Stauber:2014aa}%
  \BibitemOpen
  \bibfield  {author} {\bibinfo {author} {\bibfnamefont {T.}~\bibnamefont
  {Stauber}},\ }\href@noop {} {\bibfield  {journal} {\bibinfo  {journal} {J.
  Phys.: Condens. Matter}\ }\textbf {\bibinfo {volume} {26}},\ \bibinfo {pages}
  {123201} (\bibinfo {year} {2014})}\BibitemShut {NoStop}%
\bibitem [{\citenamefont {Brey}\ and\ \citenamefont
  {Fertig}(2007)}]{Brey:2007aa}%
  \BibitemOpen
  \bibfield  {author} {\bibinfo {author} {\bibfnamefont {L.}~\bibnamefont
  {Brey}}\ and\ \bibinfo {author} {\bibfnamefont {H.~A.}\ \bibnamefont
  {Fertig}},\ }\href {\doibase 10.1103/PhysRevB.75.125434} {\bibfield
  {journal} {\bibinfo  {journal} {Physical Review B}\ }\textbf {\bibinfo
  {volume} {75}},\ \bibinfo {pages} {125434} (\bibinfo {year}
  {2007})}\BibitemShut {NoStop}%
\bibitem [{\citenamefont {Horng}\ \emph {et~al.}(2011)\citenamefont {Horng},
  \citenamefont {Chen}, \citenamefont {Geng}, \citenamefont {Girit},
  \citenamefont {Zhang}, \citenamefont {Hao}, \citenamefont {Bechtel},
  \citenamefont {Martin}, \citenamefont {Zettl}, \citenamefont {Crommie},
  \citenamefont {Shen},\ and\ \citenamefont {Wang}}]{Horng11}%
  \BibitemOpen
  \bibfield  {author} {\bibinfo {author} {\bibfnamefont {J.}~\bibnamefont
  {Horng}}, \bibinfo {author} {\bibfnamefont {C.-F.}\ \bibnamefont {Chen}},
  \bibinfo {author} {\bibfnamefont {B.}~\bibnamefont {Geng}}, \bibinfo {author}
  {\bibfnamefont {C.}~\bibnamefont {Girit}}, \bibinfo {author} {\bibfnamefont
  {Y.}~\bibnamefont {Zhang}}, \bibinfo {author} {\bibfnamefont
  {Z.}~\bibnamefont {Hao}}, \bibinfo {author} {\bibfnamefont {H.~A.}\
  \bibnamefont {Bechtel}}, \bibinfo {author} {\bibfnamefont {M.}~\bibnamefont
  {Martin}}, \bibinfo {author} {\bibfnamefont {A.}~\bibnamefont {Zettl}},
  \bibinfo {author} {\bibfnamefont {M.~F.}\ \bibnamefont {Crommie}}, \bibinfo
  {author} {\bibfnamefont {Y.~R.}\ \bibnamefont {Shen}}, \ and\ \bibinfo
  {author} {\bibfnamefont {F.}~\bibnamefont {Wang}},\ }\href {\doibase
  10.1103/PhysRevB.83.165113} {\bibfield  {journal} {\bibinfo  {journal} {Phys.
  Rev. B}\ }\textbf {\bibinfo {volume} {83}},\ \bibinfo {pages} {165113}
  (\bibinfo {year} {2011})}\BibitemShut {NoStop}%
\bibitem [{\citenamefont {Yoon}\ \emph {et~al.}(2014)\citenamefont {Yoon},
  \citenamefont {Forsythe}, \citenamefont {Wang}, \citenamefont {Tombros},
  \citenamefont {Watanabe}, \citenamefont {Taniguchi}, \citenamefont {Hone},
  \citenamefont {Kim},\ and\ \citenamefont {Ham}}]{Yoon14}%
  \BibitemOpen
  \bibfield  {author} {\bibinfo {author} {\bibfnamefont {H.}~\bibnamefont
  {Yoon}}, \bibinfo {author} {\bibfnamefont {C.}~\bibnamefont {Forsythe}},
  \bibinfo {author} {\bibfnamefont {L.}~\bibnamefont {Wang}}, \bibinfo {author}
  {\bibfnamefont {N.}~\bibnamefont {Tombros}}, \bibinfo {author} {\bibfnamefont
  {K.}~\bibnamefont {Watanabe}}, \bibinfo {author} {\bibfnamefont
  {T.}~\bibnamefont {Taniguchi}}, \bibinfo {author} {\bibfnamefont
  {J.}~\bibnamefont {Hone}}, \bibinfo {author} {\bibfnamefont {P.}~\bibnamefont
  {Kim}}, \ and\ \bibinfo {author} {\bibfnamefont {D.}~\bibnamefont {Ham}},\
  }\href {\doibase 10.1038/nnano.2014.112} {\bibfield  {journal} {\bibinfo
  {journal} {Nature Nanotechnology}\ }\textbf {\bibinfo {volume} {9}},\
  \bibinfo {pages} {594} (\bibinfo {year} {2014})}\BibitemShut {NoStop}%
\bibitem [{\citenamefont {Peres}\ \emph {et~al.}(2013)\citenamefont {Peres},
  \citenamefont {Bludov}, \citenamefont {Ferreira},\ and\ \citenamefont
  {Vasilevskiy}}]{Peres:2013aa}%
  \BibitemOpen
  \bibfield  {author} {\bibinfo {author} {\bibfnamefont {N.~M.~R.}\
  \bibnamefont {Peres}}, \bibinfo {author} {\bibfnamefont {Y.~V.}\ \bibnamefont
  {Bludov}}, \bibinfo {author} {\bibfnamefont {A.}~\bibnamefont {Ferreira}}, \
  and\ \bibinfo {author} {\bibfnamefont {M.~I.}\ \bibnamefont {Vasilevskiy}},\
  }\href {\doibase 10.1088/0953-8984/25/12/125303} {\bibfield  {journal}
  {\bibinfo  {journal} {Journal of Physics: Condensed Matter}\ }\textbf
  {\bibinfo {volume} {25}},\ \bibinfo {pages} {125303} (\bibinfo {year}
  {2013})}\BibitemShut {NoStop}%
\bibitem [{\citenamefont {Yu}\ \emph {et~al.}(2018)\citenamefont {Yu},
  \citenamefont {Guo}, \citenamefont {Xia},\ and\ \citenamefont {Garc\'{\i}a~de
  Abajo}}]{Yu18}%
  \BibitemOpen
  \bibfield  {author} {\bibinfo {author} {\bibfnamefont {R.}~\bibnamefont
  {Yu}}, \bibinfo {author} {\bibfnamefont {Q.}~\bibnamefont {Guo}}, \bibinfo
  {author} {\bibfnamefont {F.}~\bibnamefont {Xia}}, \ and\ \bibinfo {author}
  {\bibfnamefont {F.~J.}\ \bibnamefont {Garc\'{\i}a~de Abajo}},\ }\href
  {\doibase 10.1103/PhysRevLett.121.057404} {\bibfield  {journal} {\bibinfo
  {journal} {Phys. Rev. Lett.}\ }\textbf {\bibinfo {volume} {121}},\ \bibinfo
  {pages} {057404} (\bibinfo {year} {2018})}\BibitemShut {NoStop}%
\bibitem [{\citenamefont {Ju}\ \emph {et~al.}(2011{\natexlab{b}})\citenamefont
  {Ju}, \citenamefont {Geng}, \citenamefont {Horng}, \citenamefont {Girit},
  \citenamefont {Martin}, \citenamefont {Hao}, \citenamefont {Bechtel},
  \citenamefont {Liang}, \citenamefont {Zettl}, \citenamefont {Shen},\ and\
  \citenamefont {Wang}}]{Ju:2011AA}%
  \BibitemOpen
  \bibfield  {author} {\bibinfo {author} {\bibfnamefont {L.}~\bibnamefont
  {Ju}}, \bibinfo {author} {\bibfnamefont {B.}~\bibnamefont {Geng}}, \bibinfo
  {author} {\bibfnamefont {J.}~\bibnamefont {Horng}}, \bibinfo {author}
  {\bibfnamefont {C.}~\bibnamefont {Girit}}, \bibinfo {author} {\bibfnamefont
  {M.}~\bibnamefont {Martin}}, \bibinfo {author} {\bibfnamefont
  {Z.}~\bibnamefont {Hao}}, \bibinfo {author} {\bibfnamefont {H.~A.}\
  \bibnamefont {Bechtel}}, \bibinfo {author} {\bibfnamefont {X.}~\bibnamefont
  {Liang}}, \bibinfo {author} {\bibfnamefont {A.}~\bibnamefont {Zettl}},
  \bibinfo {author} {\bibfnamefont {Y.~R.}\ \bibnamefont {Shen}}, \ and\
  \bibinfo {author} {\bibfnamefont {F.}~\bibnamefont {Wang}},\ }\href {\doibase
  10.1038/nnano.2011.146} {\bibfield  {journal} {\bibinfo  {journal} {Nature
  Nanotechnology}\ }\textbf {\bibinfo {volume} {6}},\ \bibinfo {pages} {630}
  (\bibinfo {year} {2011}{\natexlab{b}})}\BibitemShut {NoStop}%
\bibitem [{\citenamefont {Drienovsky}\ \emph {et~al.}(2018)\citenamefont
  {Drienovsky}, \citenamefont {Joachimsmeyer}, \citenamefont {Sandner},
  \citenamefont {Liu}, \citenamefont {Taniguchi}, \citenamefont {Watanabe},
  \citenamefont {Richter}, \citenamefont {Weiss},\ and\ \citenamefont
  {Eroms}}]{Drienovsky:2018aa}%
  \BibitemOpen
  \bibfield  {author} {\bibinfo {author} {\bibfnamefont {M.}~\bibnamefont
  {Drienovsky}}, \bibinfo {author} {\bibfnamefont {J.}~\bibnamefont
  {Joachimsmeyer}}, \bibinfo {author} {\bibfnamefont {A.}~\bibnamefont
  {Sandner}}, \bibinfo {author} {\bibfnamefont {M.-H.}\ \bibnamefont {Liu}},
  \bibinfo {author} {\bibfnamefont {T.}~\bibnamefont {Taniguchi}}, \bibinfo
  {author} {\bibfnamefont {K.}~\bibnamefont {Watanabe}}, \bibinfo {author}
  {\bibfnamefont {K.}~\bibnamefont {Richter}}, \bibinfo {author} {\bibfnamefont
  {D.}~\bibnamefont {Weiss}}, \ and\ \bibinfo {author} {\bibfnamefont
  {J.}~\bibnamefont {Eroms}},\ }\href {\doibase 10.1103/PhysRevLett.121.026806}
  {\bibfield  {journal} {\bibinfo  {journal} {Physical Review Letters}\
  }\textbf {\bibinfo {volume} {121}},\ \bibinfo {pages} {026806} (\bibinfo
  {year} {2018})}\BibitemShut {NoStop}%
\bibitem [{\citenamefont {Xiong}\ \emph {et~al.}(2019)\citenamefont {Xiong},
  \citenamefont {Forsythe}, \citenamefont {Jung}, \citenamefont {McLeod},
  \citenamefont {Sunku}, \citenamefont {Shao}, \citenamefont {Ni},
  \citenamefont {Sternbach}, \citenamefont {Liu}, \citenamefont {Edgar},
  \citenamefont {Mele}, \citenamefont {Fogler}, \citenamefont {Shvets},
  \citenamefont {Dean},\ and\ \citenamefont {Basov}}]{Xiong19}%
  \BibitemOpen
  \bibfield  {author} {\bibinfo {author} {\bibfnamefont {L.}~\bibnamefont
  {Xiong}}, \bibinfo {author} {\bibfnamefont {C.}~\bibnamefont {Forsythe}},
  \bibinfo {author} {\bibfnamefont {M.}~\bibnamefont {Jung}}, \bibinfo {author}
  {\bibfnamefont {A.~S.}\ \bibnamefont {McLeod}}, \bibinfo {author}
  {\bibfnamefont {S.~S.}\ \bibnamefont {Sunku}}, \bibinfo {author}
  {\bibfnamefont {Y.~M.}\ \bibnamefont {Shao}}, \bibinfo {author}
  {\bibfnamefont {G.~X.}\ \bibnamefont {Ni}}, \bibinfo {author} {\bibfnamefont
  {A.~J.}\ \bibnamefont {Sternbach}}, \bibinfo {author} {\bibfnamefont
  {S.}~\bibnamefont {Liu}}, \bibinfo {author} {\bibfnamefont {J.~H.}\
  \bibnamefont {Edgar}}, \bibinfo {author} {\bibfnamefont {E.~J.}\ \bibnamefont
  {Mele}}, \bibinfo {author} {\bibfnamefont {M.~M.}\ \bibnamefont {Fogler}},
  \bibinfo {author} {\bibfnamefont {G.}~\bibnamefont {Shvets}}, \bibinfo
  {author} {\bibfnamefont {C.~R.}\ \bibnamefont {Dean}}, \ and\ \bibinfo
  {author} {\bibfnamefont {D.~N.}\ \bibnamefont {Basov}},\ }\href
  {https://doi.org/10.1038/s41467-019-12778-2} {\bibfield  {journal} {\bibinfo
  {journal} {Nature Communications}\ }\textbf {\bibinfo {volume} {10}},\
  \bibinfo {pages} {4780} (\bibinfo {year} {2019})}\BibitemShut {NoStop}%
\bibitem [{\citenamefont {Celis}\ \emph {et~al.}(2018)\citenamefont {Celis},
  \citenamefont {Nair}, \citenamefont {Sicot}, \citenamefont {Nicolas},
  \citenamefont {Kubsky}, \citenamefont {Malterre}, \citenamefont
  {Taleb-Ibrahimi},\ and\ \citenamefont {Tejeda}}]{Celis:2018aa}%
  \BibitemOpen
  \bibfield  {author} {\bibinfo {author} {\bibfnamefont {A.}~\bibnamefont
  {Celis}}, \bibinfo {author} {\bibfnamefont {M.~N.}\ \bibnamefont {Nair}},
  \bibinfo {author} {\bibfnamefont {M.}~\bibnamefont {Sicot}}, \bibinfo
  {author} {\bibfnamefont {F.}~\bibnamefont {Nicolas}}, \bibinfo {author}
  {\bibfnamefont {S.}~\bibnamefont {Kubsky}}, \bibinfo {author} {\bibfnamefont
  {D.}~\bibnamefont {Malterre}}, \bibinfo {author} {\bibfnamefont
  {A.}~\bibnamefont {Taleb-Ibrahimi}}, \ and\ \bibinfo {author} {\bibfnamefont
  {A.}~\bibnamefont {Tejeda}},\ }\href {\doibase 10.1103/PhysRevB.97.195410}
  {\bibfield  {journal} {\bibinfo  {journal} {Physical Review B}\ }\textbf
  {\bibinfo {volume} {97}},\ \bibinfo {pages} {195410} (\bibinfo {year}
  {2018})}\BibitemShut {NoStop}%
\bibitem [{\citenamefont {Peres}\ \emph {et~al.}(2012)\citenamefont {Peres},
  \citenamefont {Ferreira}, \citenamefont {Bludov},\ and\ \citenamefont
  {Vasilevskiy}}]{Peres:2012aa}%
  \BibitemOpen
  \bibfield  {author} {\bibinfo {author} {\bibfnamefont {N.~M.~R.}\
  \bibnamefont {Peres}}, \bibinfo {author} {\bibfnamefont {A.}~\bibnamefont
  {Ferreira}}, \bibinfo {author} {\bibfnamefont {Y.~V.}\ \bibnamefont
  {Bludov}}, \ and\ \bibinfo {author} {\bibfnamefont {M.~I.}\ \bibnamefont
  {Vasilevskiy}},\ }\href@noop {} {\bibfield  {journal} {\bibinfo  {journal}
  {Journal of Physics: Condensed Matter}\ }\textbf {\bibinfo {volume} {24}},\
  \bibinfo {pages} {245303} (\bibinfo {year} {2012})}\BibitemShut {NoStop}%
\bibitem [{\citenamefont {Slipchenko}\ \emph {et~al.}(2013)\citenamefont
  {Slipchenko}, \citenamefont {Nesterov}, \citenamefont {Martin-Moreno},\ and\
  \citenamefont {Nikitin}}]{Slipchenko:2013aa}%
  \BibitemOpen
  \bibfield  {author} {\bibinfo {author} {\bibfnamefont {T.~M.}\ \bibnamefont
  {Slipchenko}}, \bibinfo {author} {\bibfnamefont {M.~L.}\ \bibnamefont
  {Nesterov}}, \bibinfo {author} {\bibfnamefont {L.}~\bibnamefont
  {Martin-Moreno}}, \ and\ \bibinfo {author} {\bibfnamefont {A.~Y.}\
  \bibnamefont {Nikitin}},\ }\href {\doibase 10.1088/2040-8978/15/11/114008}
  {\bibfield  {journal} {\bibinfo  {journal} {Journal of Optics}\ }\textbf
  {\bibinfo {volume} {15}},\ \bibinfo {pages} {114008} (\bibinfo {year}
  {2013})}\BibitemShut {NoStop}%
\bibitem [{\citenamefont {Silveiro}\ \emph {et~al.}(2013)\citenamefont
  {Silveiro}, \citenamefont {Manjavacas}, \citenamefont {Thongrattanasiri},\
  and\ \citenamefont {Garc{\'\i}a~de Abajo}}]{silveiro:2013aa}%
  \BibitemOpen
  \bibfield  {author} {\bibinfo {author} {\bibfnamefont {I.}~\bibnamefont
  {Silveiro}}, \bibinfo {author} {\bibfnamefont {A.}~\bibnamefont
  {Manjavacas}}, \bibinfo {author} {\bibfnamefont {S.}~\bibnamefont
  {Thongrattanasiri}}, \ and\ \bibinfo {author} {\bibfnamefont {F.~J.}\
  \bibnamefont {Garc{\'\i}a~de Abajo}},\ }\href {\doibase
  10.1088/1367-2630/15/3/033042} {\bibfield  {journal} {\bibinfo  {journal}
  {New Journal of Physics}\ }\textbf {\bibinfo {volume} {15}},\ \bibinfo
  {pages} {033042} (\bibinfo {year} {2013})}\BibitemShut {NoStop}%
\bibitem [{\citenamefont {Beckerleg}\ and\ \citenamefont
  {Hendry}(2016)}]{Beckerleg:2016aa}%
  \BibitemOpen
  \bibfield  {author} {\bibinfo {author} {\bibfnamefont {C.}~\bibnamefont
  {Beckerleg}}\ and\ \bibinfo {author} {\bibfnamefont {E.}~\bibnamefont
  {Hendry}},\ }\href {\doibase 10.1364/JOSAB.33.002051} {\bibfield  {journal}
  {\bibinfo  {journal} {Journal of the Optical Society of America B}\ }\textbf
  {\bibinfo {volume} {33}},\ \bibinfo {pages} {2051} (\bibinfo {year}
  {2016})}\BibitemShut {NoStop}%
\bibitem [{\citenamefont {Huidobro}\ \emph
  {et~al.}(2016{\natexlab{a}})\citenamefont {Huidobro}, \citenamefont {Kraft},
  \citenamefont {Maier},\ and\ \citenamefont {Pendry}}]{Huidobro:2016aa}%
  \BibitemOpen
  \bibfield  {author} {\bibinfo {author} {\bibfnamefont {P.~A.}\ \bibnamefont
  {Huidobro}}, \bibinfo {author} {\bibfnamefont {M.}~\bibnamefont {Kraft}},
  \bibinfo {author} {\bibfnamefont {S.~A.}\ \bibnamefont {Maier}}, \ and\
  \bibinfo {author} {\bibfnamefont {J.~B.}\ \bibnamefont {Pendry}},\ }\href
  {\doibase 10.1021/acsnano.6b01944} {\bibfield  {journal} {\bibinfo  {journal}
  {ACS Nano}\ }\textbf {\bibinfo {volume} {10}},\ \bibinfo {pages} {5499}
  (\bibinfo {year} {2016}{\natexlab{a}})}\BibitemShut {NoStop}%
\bibitem [{\citenamefont {Huidobro}\ \emph
  {et~al.}(2016{\natexlab{b}})\citenamefont {Huidobro}, \citenamefont {Kraft},
  \citenamefont {Kun}, \citenamefont {Maier},\ and\ \citenamefont
  {Pendry}}]{Huidobro:2016ab}%
  \BibitemOpen
  \bibfield  {author} {\bibinfo {author} {\bibfnamefont {P.~A.}\ \bibnamefont
  {Huidobro}}, \bibinfo {author} {\bibfnamefont {M.}~\bibnamefont {Kraft}},
  \bibinfo {author} {\bibfnamefont {R.}~\bibnamefont {Kun}}, \bibinfo {author}
  {\bibfnamefont {S.~A.}\ \bibnamefont {Maier}}, \ and\ \bibinfo {author}
  {\bibfnamefont {J.~B.}\ \bibnamefont {Pendry}},\ }\href {\doibase
  10.1088/2040-8978/18/4/044024} {\bibfield  {journal} {\bibinfo  {journal}
  {Journal of Optics}\ }\textbf {\bibinfo {volume} {18}},\ \bibinfo {pages}
  {044024} (\bibinfo {year} {2016}{\natexlab{b}})}\BibitemShut {NoStop}%
\bibitem [{\citenamefont {{Galiffi}}\ \emph {et~al.}(2019)\citenamefont
  {{Galiffi}}, \citenamefont {{Huidobro}}, \citenamefont {{Goncalves}},
  \citenamefont {{Mortensen}},\ and\ \citenamefont {{Pendry}}}]{Galiffi}%
  \BibitemOpen
  \bibfield  {author} {\bibinfo {author} {\bibfnamefont {E.}~\bibnamefont
  {{Galiffi}}}, \bibinfo {author} {\bibfnamefont {P.~A.}\ \bibnamefont
  {{Huidobro}}}, \bibinfo {author} {\bibfnamefont {P.~A.~D.}\ \bibnamefont
  {{Goncalves}}}, \bibinfo {author} {\bibfnamefont {N.~A.}\ \bibnamefont
  {{Mortensen}}}, \ and\ \bibinfo {author} {\bibfnamefont {J.~B.}\ \bibnamefont
  {{Pendry}}},\ }\href@noop {} {\bibfield  {journal} {\bibinfo  {journal}
  {arXiv e-prints}\ } (\bibinfo {year} {2019})},\ \Eprint
  {http://arxiv.org/abs/1908.09320} {arXiv:1908.09320 [cond-mat.mes-hall]}
  \BibitemShut {NoStop}%
\bibitem [{SI()}]{SI}%
  \BibitemOpen
  \href@noop {} {}\bibinfo {note} {See Supplementary Material for more details
  and additional analytical and numerical results, which includes the
  additional Refs. [40-53].}\BibitemShut {Stop}%
\bibitem [{\citenamefont {Falkovsky}\ and\ \citenamefont
  {Pershoguba}(2007)}]{Falkovsky07}%
  \BibitemOpen
  \bibfield  {author} {\bibinfo {author} {\bibfnamefont {L.~A.}\ \bibnamefont
  {Falkovsky}}\ and\ \bibinfo {author} {\bibfnamefont {S.~S.}\ \bibnamefont
  {Pershoguba}},\ }\href {\doibase 10.1103/PhysRevB.76.153410} {\bibfield
  {journal} {\bibinfo  {journal} {Phys. Rev. B}\ }\textbf {\bibinfo {volume}
  {76}},\ \bibinfo {pages} {153410} (\bibinfo {year} {2007})}\BibitemShut
  {NoStop}%
\bibitem [{\citenamefont {Stauber}\ \emph {et~al.}(2008)\citenamefont
  {Stauber}, \citenamefont {Peres},\ and\ \citenamefont {Geim}}]{Stauber08}%
  \BibitemOpen
  \bibfield  {author} {\bibinfo {author} {\bibfnamefont {T.}~\bibnamefont
  {Stauber}}, \bibinfo {author} {\bibfnamefont {N.~M.~R.}\ \bibnamefont
  {Peres}}, \ and\ \bibinfo {author} {\bibfnamefont {A.~K.}\ \bibnamefont
  {Geim}},\ }\href {\doibase 10.1103/PhysRevB.78.085432} {\bibfield  {journal}
  {\bibinfo  {journal} {Phys. Rev. B}\ }\textbf {\bibinfo {volume} {78}},\
  \bibinfo {pages} {085432} (\bibinfo {year} {2008})}\BibitemShut {NoStop}%
\bibitem [{\citenamefont {Yan}\ \emph {et~al.}(2015)\citenamefont {Yan},
  \citenamefont {Wubs},\ and\ \citenamefont {Asger~Mortensen}}]{Yan15}%
  \BibitemOpen
  \bibfield  {author} {\bibinfo {author} {\bibfnamefont {W.}~\bibnamefont
  {Yan}}, \bibinfo {author} {\bibfnamefont {M.}~\bibnamefont {Wubs}}, \ and\
  \bibinfo {author} {\bibfnamefont {N.}~\bibnamefont {Asger~Mortensen}},\
  }\href {\doibase 10.1103/PhysRevLett.115.137403} {\bibfield  {journal}
  {\bibinfo  {journal} {Phys. Rev. Lett.}\ }\textbf {\bibinfo {volume} {115}},\
  \bibinfo {pages} {137403} (\bibinfo {year} {2015})}\BibitemShut {NoStop}%
\bibitem [{\citenamefont {Zhu}\ \emph {et~al.}(2016)\citenamefont {Zhu},
  \citenamefont {Esteban}, \citenamefont {Borisov}, \citenamefont {Baumberg},
  \citenamefont {Nordlander}, \citenamefont {Lezec}, \citenamefont {Aizpurua},\
  and\ \citenamefont {Crozier}}]{Zhu16}%
  \BibitemOpen
  \bibfield  {author} {\bibinfo {author} {\bibfnamefont {W.}~\bibnamefont
  {Zhu}}, \bibinfo {author} {\bibfnamefont {R.}~\bibnamefont {Esteban}},
  \bibinfo {author} {\bibfnamefont {A.~G.}\ \bibnamefont {Borisov}}, \bibinfo
  {author} {\bibfnamefont {J.~J.}\ \bibnamefont {Baumberg}}, \bibinfo {author}
  {\bibfnamefont {P.}~\bibnamefont {Nordlander}}, \bibinfo {author}
  {\bibfnamefont {H.~J.}\ \bibnamefont {Lezec}}, \bibinfo {author}
  {\bibfnamefont {J.}~\bibnamefont {Aizpurua}}, \ and\ \bibinfo {author}
  {\bibfnamefont {K.~B.}\ \bibnamefont {Crozier}},\ }\href {\doibase
  10.1038/ncomms11495} {\bibfield  {journal} {\bibinfo  {journal} {Nature
  Communications}\ }\textbf {\bibinfo {volume} {7}},\ \bibinfo {pages} {11495}
  (\bibinfo {year} {2016})}\BibitemShut {NoStop}%
\bibitem [{\citenamefont {Park}\ \emph {et~al.}(2008)\citenamefont {Park},
  \citenamefont {Yang}, \citenamefont {Son}, \citenamefont {Cohen},\ and\
  \citenamefont {Louie}}]{Park:2008aa}%
  \BibitemOpen
  \bibfield  {author} {\bibinfo {author} {\bibfnamefont {C.-H.}\ \bibnamefont
  {Park}}, \bibinfo {author} {\bibfnamefont {L.}~\bibnamefont {Yang}}, \bibinfo
  {author} {\bibfnamefont {Y.-W.}\ \bibnamefont {Son}}, \bibinfo {author}
  {\bibfnamefont {M.~L.}\ \bibnamefont {Cohen}}, \ and\ \bibinfo {author}
  {\bibfnamefont {S.~G.}\ \bibnamefont {Louie}},\ }\href {\doibase
  10.1038/nphys890} {\bibfield  {journal} {\bibinfo  {journal} {Nature
  Physics}\ }\textbf {\bibinfo {volume} {4}},\ \bibinfo {pages} {213} (\bibinfo
  {year} {2008})}\BibitemShut {NoStop}%
\bibitem [{\citenamefont {Barbier}\ \emph {et~al.}(2008)\citenamefont
  {Barbier}, \citenamefont {Peeters}, \citenamefont {Vasilopoulos},\ and\
  \citenamefont {Pereira}}]{Barbier:2008aa}%
  \BibitemOpen
  \bibfield  {author} {\bibinfo {author} {\bibfnamefont {M.}~\bibnamefont
  {Barbier}}, \bibinfo {author} {\bibfnamefont {F.~M.}\ \bibnamefont
  {Peeters}}, \bibinfo {author} {\bibfnamefont {P.}~\bibnamefont
  {Vasilopoulos}}, \ and\ \bibinfo {author} {\bibfnamefont {J.~M.}\
  \bibnamefont {Pereira}},\ }\href {\doibase 10.1103/PhysRevB.77.115446}
  {\bibfield  {journal} {\bibinfo  {journal} {Physical Review B}\ }\textbf
  {\bibinfo {volume} {77}},\ \bibinfo {pages} {115446} (\bibinfo {year}
  {2008})}\BibitemShut {NoStop}%
\bibitem [{\citenamefont {Arovas}\ \emph {et~al.}(2010)\citenamefont {Arovas},
  \citenamefont {Brey}, \citenamefont {Fertig}, \citenamefont {Kim},\ and\
  \citenamefont {Ziegler}}]{Arovas_2010}%
  \BibitemOpen
  \bibfield  {author} {\bibinfo {author} {\bibfnamefont {D.~P.}\ \bibnamefont
  {Arovas}}, \bibinfo {author} {\bibfnamefont {L.}~\bibnamefont {Brey}},
  \bibinfo {author} {\bibfnamefont {H.~A.}\ \bibnamefont {Fertig}}, \bibinfo
  {author} {\bibfnamefont {E.-A.}\ \bibnamefont {Kim}}, \ and\ \bibinfo
  {author} {\bibfnamefont {K.}~\bibnamefont {Ziegler}},\ }\href
  {http://stacks.iop.org/1367-2630/12/i=12/a=123020} {\bibfield  {journal}
  {\bibinfo  {journal} {New Journal of Physics}\ }\textbf {\bibinfo {volume}
  {12}},\ \bibinfo {pages} {123020} (\bibinfo {year} {2010})}\BibitemShut
  {NoStop}%
\bibitem [{\citenamefont {Burset}\ \emph {et~al.}(2011)\citenamefont {Burset},
  \citenamefont {Yeyati}, \citenamefont {Brey},\ and\ \citenamefont
  {Fertig}}]{Burset_2011}%
  \BibitemOpen
  \bibfield  {author} {\bibinfo {author} {\bibfnamefont {P.}~\bibnamefont
  {Burset}}, \bibinfo {author} {\bibfnamefont {A.~L.}\ \bibnamefont {Yeyati}},
  \bibinfo {author} {\bibfnamefont {L.}~\bibnamefont {Brey}}, \ and\ \bibinfo
  {author} {\bibfnamefont {H.~A.}\ \bibnamefont {Fertig}},\ }\href
  {http://link.aps.org/doi/10.1103/PhysRevB.83.195434} {\bibfield  {journal}
  {\bibinfo  {journal} {Physical Review B}\ }\textbf {\bibinfo {volume} {83}},\
  \bibinfo {pages} {195434} (\bibinfo {year} {2011})}\BibitemShut {NoStop}%
\bibitem [{\citenamefont {Brey}\ and\ \citenamefont
  {Fertig}(2009)}]{Brey_2009}%
  \BibitemOpen
  \bibfield  {author} {\bibinfo {author} {\bibfnamefont {L.}~\bibnamefont
  {Brey}}\ and\ \bibinfo {author} {\bibfnamefont {H.~A.}\ \bibnamefont
  {Fertig}},\ }\href {\doibase 10.1103/PhysRevLett.103.046809} {\bibfield
  {journal} {\bibinfo  {journal} {Phys. Rev. Lett.}\ }\textbf {\bibinfo
  {volume} {103}},\ \bibinfo {pages} {046809} (\bibinfo {year}
  {2009})}\BibitemShut {NoStop}%
\bibitem [{\citenamefont {Vafek}(2006)}]{Vafek06}%
  \BibitemOpen
  \bibfield  {author} {\bibinfo {author} {\bibfnamefont {O.}~\bibnamefont
  {Vafek}},\ }\href {\doibase 10.1103/PhysRevLett.97.266406} {\bibfield
  {journal} {\bibinfo  {journal} {Phys. Rev. Lett.}\ }\textbf {\bibinfo
  {volume} {97}},\ \bibinfo {pages} {266406} (\bibinfo {year}
  {2006})}\BibitemShut {NoStop}%
\bibitem [{\citenamefont {G\'omez-Santos}(2009)}]{Gomez09}%
  \BibitemOpen
  \bibfield  {author} {\bibinfo {author} {\bibfnamefont {G.}~\bibnamefont
  {G\'omez-Santos}},\ }\href {\doibase 10.1103/PhysRevB.80.245424} {\bibfield
  {journal} {\bibinfo  {journal} {Phys. Rev. B}\ }\textbf {\bibinfo {volume}
  {80}},\ \bibinfo {pages} {245424} (\bibinfo {year} {2009})}\BibitemShut
  {NoStop}%
\bibitem [{\citenamefont {Stauber}\ and\ \citenamefont
  {G{\'{o}}mez-Santos}(2012)}]{Stauber12}%
  \BibitemOpen
  \bibfield  {author} {\bibinfo {author} {\bibfnamefont {T.}~\bibnamefont
  {Stauber}}\ and\ \bibinfo {author} {\bibfnamefont {G.}~\bibnamefont
  {G{\'{o}}mez-Santos}},\ }\href {\doibase 10.1088/1367-2630/14/10/105018}
  {\bibfield  {journal} {\bibinfo  {journal} {New Journal of Physics}\ }\textbf
  {\bibinfo {volume} {14}},\ \bibinfo {pages} {105018} (\bibinfo {year}
  {2012})}\BibitemShut {NoStop}%
\bibitem [{\citenamefont {Mishchenko}\ \emph {et~al.}(2010)\citenamefont
  {Mishchenko}, \citenamefont {Shytov},\ and\ \citenamefont
  {Silvestrov}}]{Mishchenko10}%
  \BibitemOpen
  \bibfield  {author} {\bibinfo {author} {\bibfnamefont {E.~G.}\ \bibnamefont
  {Mishchenko}}, \bibinfo {author} {\bibfnamefont {A.~V.}\ \bibnamefont
  {Shytov}}, \ and\ \bibinfo {author} {\bibfnamefont {P.~G.}\ \bibnamefont
  {Silvestrov}},\ }\href {\doibase 10.1103/PhysRevLett.104.156806} {\bibfield
  {journal} {\bibinfo  {journal} {Phys. Rev. Lett.}\ }\textbf {\bibinfo
  {volume} {104}},\ \bibinfo {pages} {156806} (\bibinfo {year}
  {2010})}\BibitemShut {NoStop}%
\bibitem [{\citenamefont {{Wang}}\ \emph {et~al.}(2009)\citenamefont {{Wang}},
  \citenamefont {{Nezich}}, \citenamefont {{Kong}},\ and\ \citenamefont
  {{Palacios}}}]{Wang09}%
  \BibitemOpen
  \bibfield  {author} {\bibinfo {author} {\bibfnamefont {H.}~\bibnamefont
  {{Wang}}}, \bibinfo {author} {\bibfnamefont {D.}~\bibnamefont {{Nezich}}},
  \bibinfo {author} {\bibfnamefont {J.}~\bibnamefont {{Kong}}}, \ and\ \bibinfo
  {author} {\bibfnamefont {T.}~\bibnamefont {{Palacios}}},\ }\href {\doibase
  10.1109/LED.2009.2016443} {\bibfield  {journal} {\bibinfo  {journal} {IEEE
  Electron Device Letters}\ }\textbf {\bibinfo {volume} {30}},\ \bibinfo
  {pages} {547} (\bibinfo {year} {2009})}\BibitemShut {NoStop}%
\bibitem [{\citenamefont {Stauber}\ \emph {et~al.}(2013)\citenamefont
  {Stauber}, \citenamefont {San-Jose},\ and\ \citenamefont
  {Brey}}]{Stauber:2013aa}%
  \BibitemOpen
  \bibfield  {author} {\bibinfo {author} {\bibfnamefont {T.}~\bibnamefont
  {Stauber}}, \bibinfo {author} {\bibfnamefont {P.}~\bibnamefont {San-Jose}}, \
  and\ \bibinfo {author} {\bibfnamefont {L.}~\bibnamefont {Brey}},\ }\href
  {\doibase 10.1088/1367-2630/15/11/113050} {\bibfield  {journal} {\bibinfo
  {journal} {New Journal of Physics}\ }\textbf {\bibinfo {volume} {15}},\
  \bibinfo {pages} {113050} (\bibinfo {year} {2013})}\BibitemShut {NoStop}%
\bibitem [{\citenamefont {Park}\ \emph {et~al.}(2009)\citenamefont {Park},
  \citenamefont {Son}, \citenamefont {Yang}, \citenamefont {Cohen},\ and\
  \citenamefont {Louie}}]{Park_2009}%
  \BibitemOpen
  \bibfield  {author} {\bibinfo {author} {\bibfnamefont {C.-H.}\ \bibnamefont
  {Park}}, \bibinfo {author} {\bibfnamefont {Y.-W.}\ \bibnamefont {Son}},
  \bibinfo {author} {\bibfnamefont {L.}~\bibnamefont {Yang}}, \bibinfo {author}
  {\bibfnamefont {M.~L.}\ \bibnamefont {Cohen}}, \ and\ \bibinfo {author}
  {\bibfnamefont {S.~G.}\ \bibnamefont {Louie}},\ }\href
  {http://link.aps.org/doi/10.1103/PhysRevLett.103.046808} {\bibfield
  {journal} {\bibinfo  {journal} {Physical Review Letters}\ }\textbf {\bibinfo
  {volume} {103}},\ \bibinfo {pages} {046808} (\bibinfo {year}
  {2009})}\BibitemShut {NoStop}%
\bibitem [{\citenamefont {Stauber}\ \emph {et~al.}(2017)\citenamefont
  {Stauber}, \citenamefont {G{\'o}mez-Santos},\ and\ \citenamefont
  {Brey}}]{Stauber17}%
  \BibitemOpen
  \bibfield  {author} {\bibinfo {author} {\bibfnamefont {T.}~\bibnamefont
  {Stauber}}, \bibinfo {author} {\bibfnamefont {G.}~\bibnamefont
  {G{\'o}mez-Santos}}, \ and\ \bibinfo {author} {\bibfnamefont
  {L.}~\bibnamefont {Brey}},\ }\href {\doibase 10.1021/acsphotonics.7b00524}
  {\bibfield  {journal} {\bibinfo  {journal} {ACS Photonics}\ }\textbf
  {\bibinfo {volume} {4}},\ \bibinfo {pages} {2978} (\bibinfo {year}
  {2017})}\BibitemShut {NoStop}%
\bibitem [{\citenamefont {Giuliani}\ and\ \citenamefont
  {Vignale}(2005)}]{Vignale-book}%
  \BibitemOpen
  \bibfield  {author} {\bibinfo {author} {\bibfnamefont {G.}~\bibnamefont
  {Giuliani}}\ and\ \bibinfo {author} {\bibfnamefont {G.}~\bibnamefont
  {Vignale}},\ }\href@noop {} {\emph {\bibinfo {title} {Quantum Theory of the
  Electron Liquid}}}\ (\bibinfo  {publisher} {Cambridge University Press},\
  \bibinfo {year} {2005})\BibitemShut {NoStop}%
\bibitem [{\citenamefont {A.Baldereschi}\ and\ \citenamefont
  {E.Tosatti}(1978)}]{Baldereschi-1978}%
  \BibitemOpen
  \bibfield  {author} {\bibinfo {author} {\bibnamefont {A.Baldereschi}}\ and\
  \bibinfo {author} {\bibnamefont {E.Tosatti}},\ }\href@noop {} {\bibfield
  {journal} {\bibinfo  {journal} {Solid State Communications}\ }\textbf
  {\bibinfo {volume} {29}},\ \bibinfo {pages} {131} (\bibinfo {year}
  {1978})}\BibitemShut {NoStop}%
\bibitem [{\citenamefont {Car}\ \emph {et~al.}(1981)\citenamefont {Car},
  \citenamefont {Tosatti}, \citenamefont {Baroni},\ and\ \citenamefont
  {Leelaprute}}]{Car:1981aa}%
  \BibitemOpen
  \bibfield  {author} {\bibinfo {author} {\bibfnamefont {R.}~\bibnamefont
  {Car}}, \bibinfo {author} {\bibfnamefont {E.}~\bibnamefont {Tosatti}},
  \bibinfo {author} {\bibfnamefont {S.}~\bibnamefont {Baroni}}, \ and\ \bibinfo
  {author} {\bibfnamefont {S.}~\bibnamefont {Leelaprute}},\ }\href {\doibase
  10.1103/PhysRevB.24.985} {\bibfield  {journal} {\bibinfo  {journal} {Physical
  Review B}\ }\textbf {\bibinfo {volume} {24}},\ \bibinfo {pages} {985}
  (\bibinfo {year} {1981})}\BibitemShut {NoStop}%
\bibitem [{\citenamefont {Andersen}\ \emph {et~al.}(2012)\citenamefont
  {Andersen}, \citenamefont {Jacobsen},\ and\ \citenamefont
  {Thygesen}}]{Andersen:2012aa}%
  \BibitemOpen
  \bibfield  {author} {\bibinfo {author} {\bibfnamefont {K.}~\bibnamefont
  {Andersen}}, \bibinfo {author} {\bibfnamefont {K.~W.}\ \bibnamefont
  {Jacobsen}}, \ and\ \bibinfo {author} {\bibfnamefont {K.~S.}\ \bibnamefont
  {Thygesen}},\ }\href {\doibase 10.1103/PhysRevB.86.245129} {\bibfield
  {journal} {\bibinfo  {journal} {Physical Review B}\ }\textbf {\bibinfo
  {volume} {86}},\ \bibinfo {pages} {245129} (\bibinfo {year}
  {2012})}\BibitemShut {NoStop}%
\bibitem [{\citenamefont {Wang}\ \emph {et~al.}(2015)\citenamefont {Wang},
  \citenamefont {Christensen}, \citenamefont {Jauho}, \citenamefont {Thygesen},
  \citenamefont {Wubs},\ and\ \citenamefont {Mortensen}}]{Wang:2015aa}%
  \BibitemOpen
  \bibfield  {author} {\bibinfo {author} {\bibfnamefont {W.}~\bibnamefont
  {Wang}}, \bibinfo {author} {\bibfnamefont {T.}~\bibnamefont {Christensen}},
  \bibinfo {author} {\bibfnamefont {A.-P.}\ \bibnamefont {Jauho}}, \bibinfo
  {author} {\bibfnamefont {K.~S.}\ \bibnamefont {Thygesen}}, \bibinfo {author}
  {\bibfnamefont {M.}~\bibnamefont {Wubs}}, \ and\ \bibinfo {author}
  {\bibfnamefont {N.~A.}\ \bibnamefont {Mortensen}},\ }\href
  {https://doi.org/10.1038/srep09535} {\bibfield  {journal} {\bibinfo
  {journal} {Scientific Reports}\ }\textbf {\bibinfo {volume} {5}},\ \bibinfo
  {pages} {9535 EP } (\bibinfo {year} {2015})}\BibitemShut {NoStop}%
\bibitem [{\citenamefont {Sabio}\ \emph {et~al.}(2008)\citenamefont {Sabio},
  \citenamefont {Nilsson},\ and\ \citenamefont {Castro~Neto}}]{Sabio:2008aa}%
  \BibitemOpen
  \bibfield  {author} {\bibinfo {author} {\bibfnamefont {J.}~\bibnamefont
  {Sabio}}, \bibinfo {author} {\bibfnamefont {J.}~\bibnamefont {Nilsson}}, \
  and\ \bibinfo {author} {\bibfnamefont {A.~H.}\ \bibnamefont {Castro~Neto}},\
  }\href {\doibase 10.1103/PhysRevB.78.075410} {\bibfield  {journal} {\bibinfo
  {journal} {Physical Review B}\ }\textbf {\bibinfo {volume} {78}},\ \bibinfo
  {pages} {075410} (\bibinfo {year} {2008})}\BibitemShut {NoStop}%
\bibitem [{\citenamefont {Strait}\ \emph {et~al.}(2013)\citenamefont {Strait},
  \citenamefont {Nene}, \citenamefont {Chan}, \citenamefont {Manolatou},
  \citenamefont {Tiwari}, \citenamefont {Rana}, \citenamefont {Kevek},\ and\
  \citenamefont {McEuen}}]{Strait13}%
  \BibitemOpen
  \bibfield  {author} {\bibinfo {author} {\bibfnamefont {J.~H.}\ \bibnamefont
  {Strait}}, \bibinfo {author} {\bibfnamefont {P.}~\bibnamefont {Nene}},
  \bibinfo {author} {\bibfnamefont {W.-M.}\ \bibnamefont {Chan}}, \bibinfo
  {author} {\bibfnamefont {C.}~\bibnamefont {Manolatou}}, \bibinfo {author}
  {\bibfnamefont {S.}~\bibnamefont {Tiwari}}, \bibinfo {author} {\bibfnamefont
  {F.}~\bibnamefont {Rana}}, \bibinfo {author} {\bibfnamefont {J.~W.}\
  \bibnamefont {Kevek}}, \ and\ \bibinfo {author} {\bibfnamefont {P.~L.}\
  \bibnamefont {McEuen}},\ }\href {\doibase 10.1103/PhysRevB.87.241410}
  {\bibfield  {journal} {\bibinfo  {journal} {Phys. Rev. B}\ }\textbf {\bibinfo
  {volume} {87}},\ \bibinfo {pages} {241410} (\bibinfo {year}
  {2013})}\BibitemShut {NoStop}%
\end{thebibliography}%
\vspace{0.5truecm}
\end{document}